\documentclass[fleqn,usenatbib]{mnras}

\usepackage{newtxtext,newtxmath}
\usepackage{diagbox}
\usepackage{float}
\usepackage{stfloats}
\usepackage{scalerel}
\usepackage{stackengine,wasysym}
\usepackage{breqn}

\usepackage[T1]{fontenc}

\usepackage{multirow}

\DeclareRobustCommand{\VAN}[3]{#2}
\let\VANthebibliography\thebibliography
\def\thebibliography{\DeclareRobustCommand{\VAN}[3]{##3}\VANthebibliography}

\usepackage{graphicx}	
\usepackage{amsmath}	
\usepackage{amssymb}	

\usepackage{xargs}
\usepackage[pdftex,dvipsnames]{xcolor}
\newcommand{\Quijote}{\textsc{Quijote} }
\newcommand{\Mpc}{\rm{Mpc}}
\newcommand{\hMpc}{h^{-1}\ \Mpc}
\newcommand{\hMsun}{h^{-1}\ M_{\odot}}
\newcommand{\de}{\text{d}}

\title[Reconstruction analysis]{Analysis of an iterative reconstruction method in comparison of the standard reconstruction method}

\author[Chen \& Padmanabhan]{
Xinyi Chen,$^{1}$\thanks{E-mail: xinyi.chen@yale.edu}
Nikhil Padmanabhan$^{1,2}$
\\
$^{1}$Department of Physics, Yale University, New Haven, CT, USA\\
$^{2}$Department of Astronomy, Yale University, New Haven, CT, USA\\
}

\date{Accepted XXX. Received YYY; in original form ZZZ}

\pubyear{2023}

\begin{document}
\label{firstpage}
\pagerange{\pageref{firstpage}--\pageref{lastpage}}
\maketitle

\begin{abstract}
We present a detailed analysis of a new, iterative density reconstruction algorithm. This algorithm uses a decreasing smoothing scale to better
reconstruct the density field in Lagrangian space. We implement this algorithm to 
run on the \Quijote simulations, and extend it to (a) include a smoothing kernel 
that smoothly goes from anisotropic to isotropic, and (b) a variant that does not
correct for redshift space distortions. We compare the performance of this
algorithm 
with the standard reconstruction method. 
Our examinations of the methods include cross-correlation
of the reconstructed density field with the linear density field, reconstructed
two-point functions, and BAO parameter fitting.
We also examine the impact of various parameters, such as smoothing scale, anisotropic smoothing, 
tracer type/bias, and the inclusion of second order perturbation theory. 
We find that the two reconstruction algorithms are
comparable in most of the areas we examine. In particular, both algorithms give
consistent fittings of BAO parameters. The fits are robust over a range of
smoothing scales. We find the iterative algorithm is significantly better at
removing redshift space distortions. 
The new algorithm
will be a promising method to be employed in the ongoing and future large-scale
structure surveys.

\end{abstract}

\begin{keywords}
cosmology: large-scale structure of Universe -- methods: numerical -- methods: statistical
\end{keywords}



\section{Introduction}

The baryon acoustic oscillation (BAO) technique has become one of the most
promising probes of dark energy over the last decade. BAO imprints a preferred
scale in the clustering of matter and galaxies, known as the ``acoustic scale'',
which can be used as a standard ruler to map the expansion history of the
Universe and study dark energy. This technique relies on precise measurements of
the acoustic peak in configuration space, or equivalently, the wiggles in
Fourier space. Ongoing and future ground-based galaxy surveys, such as DESI
\citep{DESI} and Vera Rubin Observatory/LSST \citep{LSST}, and space-based
surveys, such as \textit{Euclid} \citep{Laureijs11} and \textit{Nancy Roman
Space Telescope} \citep{Spergel13}, will achieve unprecedented precision of
measurements of dark energy and other cosmological parameters, thus shedding
light on the nature of dark energy.   

One of the systematics that impedes the precise measurements of BAO arises from
the nonlinear evolution of the density field in the late-time universe,
manifested in e.g., large-scale bulk flows and supercluster formation. This
nonlinear evolution partially erases the acoustic peak in the matter correlation
function and damps the higher harmonics in the power spectrum, reducing the
accuracy of the BAO distance measurement
\citep[e.g.,][]{Meiksin99,Seo05,Eisenstein07b}. Density field reconstruction
aims to mitigate this effect by reversing the large-scale bulk flows. The
standard reconstruction method uses Lagrangian perturbation theory to  estimate
the displacement of galaxies caused by nonlinear evolution and moves particles
back by this distance \citep[][]{Eisenstein07}. This method has been applied to
data \citep[
e.g.,][]{Padmanabhan12,Anderson12,Xu13,Anderson14b,Kazin14,Ross15,Beutler15,Alam17,Gil20}
in the past decade and has achieved improvements of precision by about a factor
of 1.2-2.4 \citep[e.g.,][]{Padmanabhan12,Xu13,Anderson14b,Gil20}. However, next
generation surveys could benefit from further improvement to fully realize their
potential. Towards this end, there have been a number of new reconstruction
methods proposed recently, aiming to obtain better results than the standard
method.

Reconstruction methods can generally be categorized as Lagrangian or Eulerian,
depending on whether or not moving particles back is part of the procedure. An
incomplete list of Lagrangian algorithms includes the first method applied to
BAO surveys, which was henceforth referred to as the standard method
\citep{Eisenstein07}, iterative methods \citep[e.g.,][]{Seo10,Tassev12},
iterative methods with annealing smoothing scales
\citep[e.g.,][]{Schmittfull17}, and Lagrangian methods with pixels
\citep[e.g.,][]{Seo16,Obuljen17}. An incomplete list of Eulerian algorithms
involves iterative methods with annealing smoothing scales \citep{Hada18} and
Eulerian with growth-shift \citep{Schmittfull15}. Among the various new
ingredients, iterative approaches
\citep[e.g.,][]{Seo10,Tassev12,Schmittfull17,Hada18} aim to better resolve the
difference between the Lagrangian and Eulerian positions, and varying smoothing
\citep[e.g.,][]{Schmittfull17,Hada18} is designed to access smaller scales. In
addition, methods may use second-order perturbation theory
\citep[e.g.,][]{Hada18} to attempt to reduce the inaccuracies of the
approximation.

We choose \citet[][hereafter, HE18]{Hada18} as representative of the new
techniques to conduct detailed comparisons with the standard method
\citep[][hereafter, ES3]{Eisenstein07}. The HE18 algorithm does not move
particles back, but maps the density field instead. In observations, moving
particles might cause mismatches across survey boundaries; thus, restoring the
density field without moving particles may be a better approach. Also, it is an
iterative method with decreasing (annealing) smoothing scale in each iteration.
Additionally, HE18 uses the second-order perturbation theory to estimate the
displacement field. However, both HE18 and ES3 share common steps, such as
estimating the displacement fields.

A reconstruction algorithm can be analyzed via cross-correlation with linear
density field, power spectrum and BAO fitting ability. A better reconstruction
algorithm should produce a reconstructed field closer to the linear density
field, a power spectrum better resembles the linear power spectrum, and
resultant lower errors in BAO distance measurements. Our statistic metrics
include propagator and cross-correlation of density fields, monopole power
spectrum of reconstructed fields, quadrupole power spectrum in redshift space,
and fittings of BAO parameters. We use \Quijote simulations \citep{Navarro19} to
carry out detailed analysis that scrutinizes these aspects.  We analyze the
dependence of the performance on parameters, such as isotropic and anisotropic
smoothing scale, number density of particles, and galaxy bias. 

The structure of this paper is as follows. Section~\ref{sect:methods} summarizes
the two algorithms. Section~\ref{sect:simulations} presents the simulations we
use. Sections~\ref{sect:field} and ~\ref{sect:two-point} present analysis
regarding reconstructed fields and two-point functions.
Section~\ref{sec:BAO-fit} analyzes BAO fitting results.
Sections~\ref{sec:discussion} and~\ref{sec:concl} are discussions and
conclusions. Throughout this paper, we use \Quijote's fiducial cosmology, which
is close to Planck 2018 \citep[][]{Planck18}. We use ${\bf s}$ to denote Eulerian
positions and ${\bf q}$ for Lagrangian positions. The subscript $s$ denotes for
redshift space.

\section{Overview of reconstruction methods}\label{sect:methods}

We review the two reconstruction algorithms we analyze in this study in this
section. Reconstruction can be done with or without removing the redshift space
distortions (RSD) with the corresponding isotropic and anisotropic
reconstruction conventions. We summarize first the standard algorithm, i.e. ES3
algorithm, and the new iterative algorithm, i.e. HE18, with isotropic
reconstruction convention. For HE18, we also explore anisotropic smoothing,
which we describe following the introduction of HE18 algorithm. We analyze
anisotropic reconstruction for the two algorithms as well, and we describe the
approaches in the end of this section.

\subsection{ES3 reconstruction algorithm (isotropic)}

The ES3 reconstruction method \citep[][]{Eisenstein07} aims to reconstruct the acoustic peak by moving
particles back by their Zel'dovich displacements \citep[][]{Zeldovich70}. 
We summarize their procedure (as implemented in this paper) below.

(1) Distribute particles on a grid following Triangular Shaped Cloud (TSC) method. Calculate the density field of this grid, $\delta({\bf k})$. 

(2) Compute the Zel'dovich displacement (the first order solution in
Euler-Poisson systems of equations from perturbation theory)
\citep[e.g.,][]{Zeldovich70,Buchert93,Buchert94}, for the smoothed density
field: 

\begin{equation}\label{eqn:standard_dis}
{\bf S(k)}_{s}={\bf S(k)}+f[{\bf S(k)}\cdot {\hat{\bf z}}]{\hat {\bf z}},\\
{\bf S}({\bf k})=\frac{i({\bf k}/k^2)\delta({\bf k})G(k)}{b(1+\beta\mu^2)},  
\end{equation}
in redshift space for periodic boxes. $\delta({\bf k})$ is the density field,
$b$ is the linear galaxy bias, and $\beta=f/b$, where $f$ is the linear growth
rate. In redshift space, the Zel'dovich displacement in the line-of-sight
direction contains the Kaiser factor, where $f$ is again the estimated linear
growth rate and is set as an input parameter to study the impact of its
mis-estimation. 
The density field is divided by bias as well as the Kaiser
factor when constructed from particles. The last term, $G(k)$, is the smoothing
kernel, usually assumed to be a Gaussian of width $R$: 
\begin{equation}\label{eqn:smoothing}
G(k)=\exp\left(-\frac{1}{2}k^2R^2\right).
\end{equation}
We smooth the field because small-scale data is noisy and less accurate,
especially before reconstruction. The real space case is
equation~\ref{eqn:standard_dis} when $f=0$.

(3) Move the original particles by $-{\bf S}({\bf k})_{s}$ and obtain the ``displaced'' density field, $\delta_{\rm displaced}$.

(4) Shift a set of randomly distributed particles over the same geometry. In
both redshift and real space, move these randomly distributed particles by $-{\bf
S}({\bf k})$ in equation~\ref{eqn:standard_dis} to form the ``shifted'' density
field, $\delta_{\rm shifted}$. This step restores the power at large scales (low-$k$) that would otherwise be lost when shifting particles back to their original positions. Here we adopt the isotropic reconstruction
convention, where we attempt to remove the RSD by
moving the particles and the randoms by different distances (c.f. anisotropic
reconstruction in Section~\ref{sec:RecAni_explain}).

(5) Compute the difference of the two fields to obtain the reconstructed density field: $\delta_r=\delta_{\rm displaced}-\delta_{\rm shifted}$.

\subsection{HE18 reconstruction algorithm (isotropic)}
The HE18 method \citep[][]{Hada18} does not move particles, but instead it advects the density
field. It approximates the linear density field as two parts in Lagrangian
space, a large-scale part, $\delta_l({\bf q},t)$, and a small-scale residual
part, $\delta_{\rm res}({\bf q})$: $\delta_L({\bf q},t)=\delta_l({\bf
q},t)+\delta_{\rm res}({\bf q})$. Only the large-scale part, $\delta_l$, is
assumed to cause displacement and thus changes over time, and the residual part
exists from the beginning and stays the same over time. Hence, it modifies the
assumption made in Lagrangian perturbation theory that the initial density everywhere is the mean density,
$\rho({\bf q})={\bar{\rho}}$; instead, $\rho({\bf q})=(1+\delta_{\rm res}({\bf
q})){\bar \rho}$. This then changes the Jacobian of the transformation between
Eulerian and Lagrangian space, which is now
\begin{equation}
\det[\delta_{ab}^K+S_{a,b}]=\frac{\rho({\bf q})}{\rho({\bf s})}=\frac{1+\delta_{\rm res}({\bf q})}{1+\delta({\bf s})}.
\end{equation}

Furthermore, with the assumption that only the large-scale part causes displacement, the linear solution to the Euler-Poisson system of equations becomes 
\begin{equation}
\nabla \cdot {{\bf S}^{(1)}({\bf q})}=\nabla \cdot {{\bf S}_{l}^{(1)}({\bf q})}=-\delta_{l}({\bf q})=-\delta_{L}({\bf q})G(k), 
\end{equation}
where ${\bf S}_{l}^{(1)}$ is the first-order displacement of the large-scale part of the density field. The smoothed linear density is considered as the large-scale density. $G(k)$ is the smoothing kernel, equation~\ref{eqn:smoothing}, with isotropic smoothing (c.f. section~\ref{sec:anisotropic_smoothing} for anisotropic smoothing). $R$ is the smoothing scale, but it has the annealing feature in HE18, i.e., decreasing over iterations (see below), which is different from ES3's one-step, unchanged smoothing.

The method can be described as follows:

(1) Distribute particles on a grid following TSC method. Calculate the density
field of this grid $\delta_s({\bf s})$. This step is the same as the first step
in ES3 method. 

(2) Starting with the above calculated density in Eulerian space as the linear
density in Lagrangian space, $\delta_L({\bf q})=\delta_s({\bf s})$ (we move to
the Lagrangian space here before iterations and remain in the Lagrangian space
throughout the iterations), estimate the first and second order displacements
${\bf S}_l^{(1)}$, ${\bf S}_l^{(2)}$, and ${\bf S}_l^{(s)}$ in the following:
\begin{equation}\label{eqn:Sl1}
{\bf S} _l^{(1)}({\bf k})=\frac{i {\bf k}}{k^2}\delta_L({\bf k})G(k),
\end{equation}
and
\begin{equation}\label{eqn:S2}
{\bf S}_l^{(2)}({\bf k})=-\frac{i{\bf k}}{k^2}{\rm FFT}
\left[
-\frac{3}{14}\Omega_m^{-1/143}\sum_{a\neq b}
   \left(S_{a,a}^{(1)}S_{b,b}^{(1)}-S_{a,b}^{(1)}S_{b,a}^{(1)}\right)
\right]
\end{equation}
where $S_{a,b}=\partial S_{a}/\partial q_{b}$, and $a$ and $b$ run over the Cartesian coordinates. Here, ${\bf S}_{l}^{(1)}$ and
${\bf S}_{l}^{(2)}$ estimate the ${\bf S}^{(1)}$ and ${\bf S}^{(2)}$, the first
(or Zel'dovich displacement) and second orders of solutions to the Euler-Poisson system of
equations, by the aforementioned assumption. Iterations start with an ${\bf
S}^{(1)}$ that is the same as ${\bf S}({\bf k})$ in the ES3 method in
equation~\ref{eqn:standard_dis}, except that only the bias $b$, but not the
Kaiser factor, is being corrected for the density, i.e. $\delta_L({\bf
k})=\delta({\bf k})/b$. Redshift space distortions are being accounted for
through iterative estimations of the displacement. Although
equation~\ref{eqn:Sl1} is only valid in linear theory and our pre-reconstruction
density contains RSD, the iterations will converge regardless of where the input
position is.

In order to better access smaller scales, we decrease the smoothing scale using the prescription in HE18 and \citet[][]{Schmittfull17}:
\begin{equation}
    R_{n}={\rm max}\left(\frac{R_{\rm ini}}{\mathcal{D}^n},R_{\rm eff}\right),
\end{equation}
where subscript $n$ denotes iteration number. $R_{\rm ini}$ is the smoothing scale we start iterations with and we use $R_{\rm ini}=20\ \hMpc$ in all cases. $\mathcal{D}$ is the annealing factor. We use 1.2 for our tests. 
If the smoothing scale reaches the effective smoothing and the iteration has not converged, the algorithm runs at the effective smoothing, $R_{\rm eff}$, for the remaining iterations. Thus, $R_{\rm eff}$ is the final smoothing scale, unless the iterations converge before reaching that value.

With ${\bf S}_{l}^{(1)}$ and ${\bf S}_{l}^{(2)}$ calculated, we can obtain the total displacement:
\begin{equation}\label{eqn:S_l_real}
{\bf S}_{l}({\bf q},t)={\bf S}_{l}^{(1)}+{\bf S}_{l}^{(2)}
\end{equation}
in real space and 
\begin{equation}~\label{eqn:S_s}
{\bf S}_{l,s}({\bf q},t)={\bf S}_{l}^{(1)}+{\bf S}_{l}^{(2)}+\beta[({\bf S}_{l}^{(1)}+2{\bf S}_{l}^{(2)})\cdot {\hat{{\bf z}}}]{\bf {\hat z}} 
\end{equation}
in redshift space.

(3) Update the linear density following 
\begin{equation}
\begin{split}
\delta_{L,n,{\rm pre-weighted}}({\bf q},t)&=\det \left[\delta_{ab}^{K}+S_{l,s,n|a,b}\right]\times \left(1+\delta_{s,n}({\bf s[{\bf q}]})
\right)\\
& -\left(1+\mu_{1}({\bf S}_{l,n-1}^{(1)}({\bf q}))\right) 
\end{split}
\end{equation}
through iterations until the linear density field converges. The subscript $n$ is again the iteration number, while the subscript $s$ denotes redshift space. The vertical line is to separate the notation from derivative, i.e. $S_{l,s,n|a,b}$ is the derivative of the $S_{a}$ component with respect to coordinate $q_b$ (see Eqn.~\ref{eqn:S2}) at the $n$'th iteration. Again, $a$ and $b$ run over the Cartesian coordinates. Here, 
\begin{equation}
\mu_1({\bf S}_{l}^{(1)})=\nabla \cdot {\bf S}_{l}^{(1)}=-\delta_l({\bf q}).
\end{equation}
In real space, one needs to substitute ${\bf S}_{l|a,b}$ for ${\bf S}_{l,s|a,b}$. Note that the method starts with final Eulerian particle position {\bf s}, but the estimated linear density field is in terms of the initial, Lagrangian particle position {\bf q}. $\delta_s({\bf s})$ is unsmoothed, while $\mu_1({\bf S}_{l}^{(1)})$ and ${\bf S}_{l,s|a,b}$ (or ${\bf S}_{l|a,b}$) are both smoothed. One place we modify from HE18 original algorithm is that we calculate the above determinant directly, instead of expressing it in terms of the derivatives of ${\bf S}_l^{(s)}$. 

(4) Weigh the calculated $n$-th density field:
\begin{equation}
    \delta_{L,n}=w\delta_{L,n,{\rm pre-weighted}}+(1-w)\delta_{L,n-1},
\end{equation}
where $w$ is the weight, which takes a value between 0 and 1. $w$ varies according to the smoothing scale. 
The particular choices of weights that we use (corresponding to different smoothing scales) is specified below.

The convergence criterion is 
\begin{equation}
r_{\rm con}=\frac{\sum [\delta_{L,n}-\delta_{L,n-1}]^2}{\sum \delta_s^2}<0.01,
\end{equation}
where the subscript $n$ represents the iteration number and the summation is over all the grid cells. $\delta_s$ is the unsmoothed pre-reconstruction density.

\subsubsection{Anisotropic smoothing}\label{sec:anisotropic_smoothing}
The smoothing described above is isotropic. However, due to RSD along line of sight, we might benefit from an anisotropic smoothing. Following HE18, we introduce a parameter that quantifies the different smoothings along line of sight and perpendicular to line of sight: 
\begin{equation}
C_{\rm ani}=R_{\parallel}/R_{\perp},
\end{equation}
where $R_{\parallel}$ and $R_{\perp}$ represent the smoothing scale in the direction parallel and perpendicular to line of sight, respectively. Accordingly, the anisotropic smoothing becomes
\begin{equation}
G_{\rm ani}(k)=\exp[-\frac{1}{2}(k_{\perp}^2+k_{\parallel}^2C_{\rm ani}^2)R_{\perp}^2].    
\end{equation}
The effective smoothing in the isotropic case corresponds to the perpendicular to line-of-sight smoothing, $R_{\perp}$. We analyze two cases for HE18, one with fixed $C_{\rm ani}$ over iterations, and the other with annealing $C_{\rm ani}$, which decreases by a factor of 1.1 every iteration until it becomes 1:
\begin{equation}
    C_{{\rm ani},n}={\rm max}\left(\frac{C_{\rm ani,ini}}{1.1^n},1\right).
\end{equation}
We attempt to reduce the smoothing needed along the line of sight, because the effects of RSD decrease over iterations where RSD is gradually being removed. 

We analyze both algorithms with smoothing scales 
15, 10, and 7.5 $\hMpc$ for isotropic smoothing. For anisotropic smoothing analysis, $R_{\perp}$ are fixed at these values and $C_{\rm ani}$ (or $C_{\rm ani,ini}$, in the case of annealing $C_{\rm ani}$) varies from 1 to 2.5. The weight $w$ is taken to be 0.7, 0.5, and 0.4, corresponding to the above smoothing scales.

\subsection{Anisotropic reconstruction}\label{sec:RecAni_explain}

Our regular reconstruction (with isotropic and anisotropic smoothing) in redshift space is isotropic, meaning that we attempt to remove redshift space distortions in the final reconstructed field. Since reconstruction does not completely remove RSD, we explore an alternative reconstruction approach where we do not remove RSD and can thus better model the output. We examine the performance of the two algorithms in redshift space when RSD is not removed. 

For HE18, we use the same pre-reconstruction density field (not divided by the Kaiser factor) to calculate the displacement field. We calculate the displacement without the redshift space part in equation~\ref{eqn:S_s}:
\begin{equation}
    {\bf S}_{l}({\bf q},t)={\bf S}_{l}^{(1)}+{\bf S}_{l}^{(2)},
\end{equation}
effectively the same as equation~\ref{eqn:S_l_real} for real space. For ES3, we also use the same pre-reconstruction density field, which is divided by the Kaiser factor. We move both the displaced and the shifted density fields by the same amount,
\begin{equation}
{\bf S(k)}_{s}={\bf S(k)}+f[{\bf S(k)}\cdot {\hat{\bf z}}]{\hat {\bf z}}.
\end{equation}
This is one convention for anisotropic reconstruction \citep[e.g.,][denoted as RecSym]{Seo08,Chen19} (c.f. different conventions in \citet[][]{Seo10,Mehta11,Seo16b}). 

\section{Overview of simulations used}\label{sect:simulations}

We use the \textsc{Quijote} simulations \citep{Navarro19} to conduct our
analysis. The \textsc{Quijote} simulations are a large suite of full $N$-body
simulations run with TreePM Gadget-III for three different resolutions in boxes
of 1 $h^{-1}$Gpc using a cosmology close to Planck 2018 cosmology
\citep{Planck18}: $\Omega_{\rm m}=0.3175$, $\Omega_{\rm b}=0.049$, $h=0.6711$, $n_s=0.9624$, $\sigma_8=0.834$, $M_{\nu}=0.0\ $eV, and $w=-1$. 
We use the 100
high-resolution (1024$^3$ CDM particles) simulations with both snapshots as
matter fields and halo catalogs as halo fields at redshift $z=0$ and 1.
We focus on two mass bins for the halo fields, $10^{12.5}-10^{13}
\hMsun$ and $10^{13}-10^{13.5} \hMsun$. The number densities for the two mass
bins are $1.0\times10^{-3} \hMpc^3$ and $3.7\times10^{-4} \hMpc^3$ at $z=0$, and
$8.6\times10^{-4} \hMpc^3$ and $2.3\times10^{-4} \hMpc^3$ at $z=1$, respectively.
We estimate the bias for the two mass bins to be 0.92 and 1.12 at $z=0$, and 1.70
and 2.29 at $z=1$, respectively.

\section{Comparisons of the reconstructed fields}

As described above, the reconstruction algorithms have a number of parameters that 
can be tuned. In this section, we examine the performance of the algorithms as function of these
parameters at the level of the reconstructed density fields. While we will compare the 
ES3 and HE18 algorithms as part of this, we defer a more detailed comparison to 
Sec.~\ref{sec:he18_vs_es3}.

\subsection{Propagators and cross-correlations}\label{sect:field}

\begin{figure}
    \centering
    \includegraphics[width=\columnwidth]{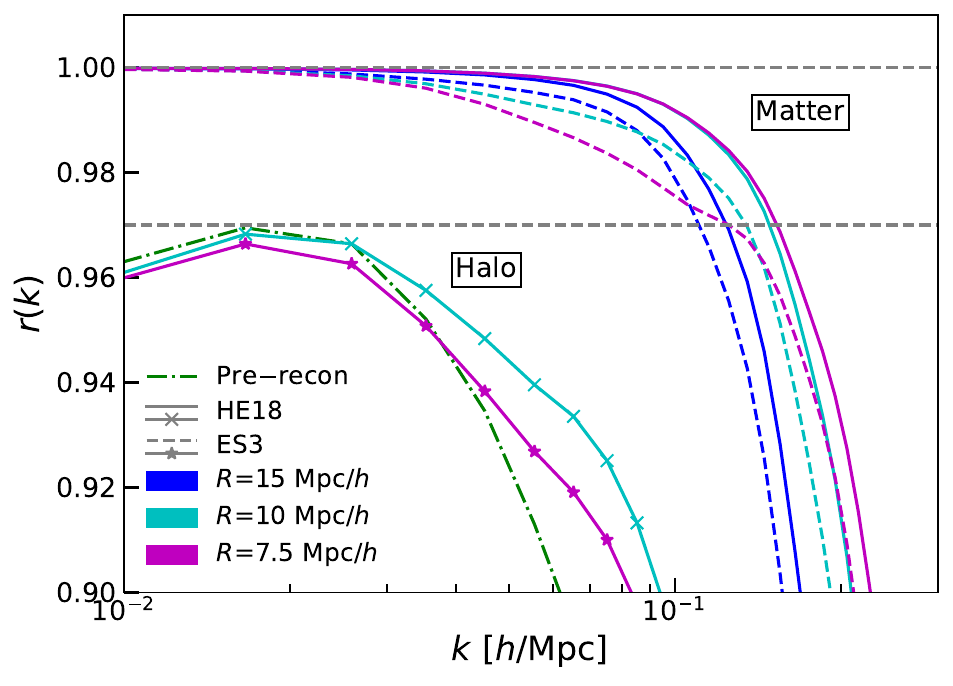}
    \caption{The cross-correlation $r(k)$ with the initial density field for
    lower-mass bin halo and matter fields at $z=0$ in redshift space. Solid
    lines are HE18 matter field, dashed lines are ES3 matter field, crosses are
    HE18 halo field, stars are ES3 halo field. The colors - blue, cyan and magenta represent 15, 10,
    and 7.5 $\hMpc$ smoothing. 
    The horizontal dashed lines are the expected 
    $r(k)$ values as $k \rightarrow 0$. For halos, this value is less than 1 due 
    to shot noise. 
    In general, we find that decreasing the smoothing scale improves the overall 
    cross-correlation with the initial density field. However, this improvement appears to
    reach a maximum at $\sim 10 \hMpc$ for the ES3 algorithm, after which the cross-correlation
    coefficient starts to get distorted. The HE18 algorithm does not appear to show this 
    behavior. We find similar trends with smoothing for the halo fields, 
    although we do not see any distortions in $r(k)$ in the ES3 algorithm 
    at the smoothing scales we consider.
    We only show the 
    results for the $7.5 \hMpc$ smoothing scale for ES3, and the $10 \hMpc$ smoothing 
    scale for HE18 for clarity. As a reference, the dash-dotted line shows the cross-correlation 
    with the unreconstructed halo field.}
    \label{fig:halo12513_z0_zspace_matter_z0_zspace_rk}
\end{figure}

The goal of reconstruction is to recover the linear density field, and so, the 
simplest (two-point) statistic to examine is the cross-correlation between the 
reconstructed and linear density fields. There are two ways to normalize this 
cross-correlation. The first is the standard cross-correlation coefficient $r(k)$
defined as 
\begin{equation}
r(k)=\frac{\left<\delta^{*}(k)\delta_{\rm ini}(k)\right>}{\sqrt{\left<\delta^2(k)\right>\left<\delta^2_{\rm ini}(k)\right>}}
\end{equation}
where
$\delta(k)$ and $\delta_{\rm ini}(k)$ are the reconstructed and the initial densities in Fourier space, respectively. 
Note that $r(k)$ is strictly bounded between $-1$ and $1$, 
and the closer it is to $1$, the better the reconstruction.
However, $r(k)$ is not sensitive to the overall ($k$-dependent) amplitude of the reconstructed field, but it measures
how well reconstruction has recovered the phases of the initial density field.

A second normalization is the propagator $G(k)$, which is defined as
\begin{equation}
G(k)=\frac{\left<\delta^{*}(k)\delta_{\rm ini}(k)\right>}{\left<\delta^2_{\rm ini}(k)\right>} 
\end{equation}
This is commonly used in studies of reconstructing the BAO feature since it characterizes the 
nonlinear damping of the BAO feature. Initial studies of reconstruction \citep{Eisenstein07b,Padmanabhan09b}
modeled this as a Gaussian. A more recent study \citep{Seo16b} showed that a modification
to a simple Gaussian form can better account for the impact of the reconstruction smoothing scale.
They propose a form for the propagator 
\begin{equation}\label{eqn:modified_gaussian}
    G(k,\mu)=(1+\beta\mu^2\Sigma_R)
       \exp\left[-\frac{1}{4}
       \left(k^2\mu^2\Sigma_{\parallel}^2+k^2(1-\mu^2)\Sigma_{\perp}^2\right)\right],\\
\end{equation}
where $\Sigma_R=1$ before reconstruction and
\begin{equation}
    \Sigma_R=1-\exp[-k^2R^2/2]
\end{equation}
after reconstruction. The parameters $\Sigma_{\parallel}$ and $\Sigma_{\perp}$ describe
the BAO damping in the line-of-sight and transverse directions, respectively, while 
$R$ is related to the reconstruction smoothing scale.  
Measuring the propagator and these three 
parameters allow us to quantify the impact of reconstruction on the BAO feature. 
To avoid fitting to the full 2d shape of the propagator, we fit the $l=0,2$ multipoles
defined by
\begin{equation}\label{eqn:G_l}
    G_l(k)=\frac{2l+1}{2}\int_{-1}^{1}G(k,\mu) L_l(\mu)\de \mu \,,
\end{equation}
where $L_l(\mu)$ is the Legendre polynomial. 

Finally, we note that the ratio of $G(k)$ to $r(k)$ is the square-root of the 
ratio of the reconstructed power spectrum to the linear power spectrum as well. 

\begin{figure}
    \centering
    \includegraphics[width=\columnwidth]{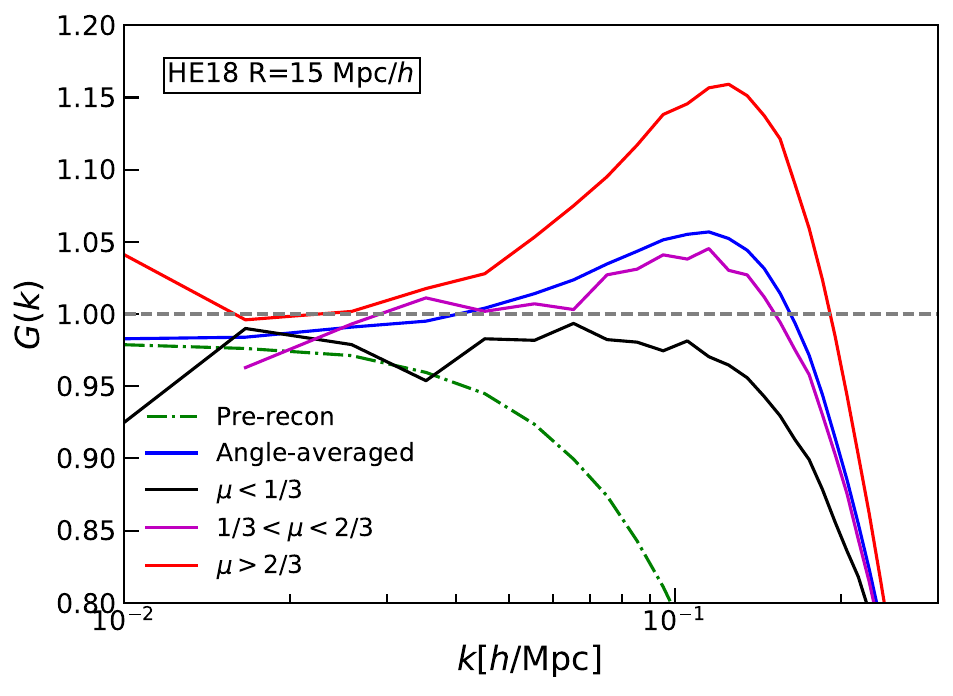}
    \caption{The propagator $G(k)$ as a function of angle, shown here for the 
    lower mass halo field in redshift space. The field is 
    reconstructed using HE18 with a 15 $\hMpc$ effective smoothing scale, with
    qualitatively similar results for other smoothing scales, as well as the ES3 algorithm.
    The black, magenta and red lines correspond to angular wedges of 
    $\mu < 1/3$ (perpendicular to line-of-sight), $1/3 < \mu < 2/3$, and 
    $\mu > 2/3$ (parallel to line-of-sight), respectively, where $\mu = |k_z/k|$.
    The blue line shows the angle-averaged $G(k)$. The green dash-dotted line is the 
    propagator before reconstruction (in redshift space) for comparison. 
    The angle groups show some noise, but we observe anti-correlations among them.
    The shape of the propagator and in particular, the bump
    around $k=0.1\hMpc$ can clearly be seen as coming from redshift space distortions,
    and this shape is captured by the modified Gaussian model presented in the text.
    } 
    \label{fig:halo12513_z0_zspace_gk_gkangle}
\end{figure}

We start by measuring $r(k)$ and $G(k)$ as a function of the reconstruction 
smoothing scale. Fig.~\ref{fig:halo12513_z0_zspace_matter_z0_zspace_rk} 
plots $r(k)$ for different smoothing scales for different algorithms, for 
both the matter and halo density fields. In general, we find that the cross-correlation
improves with decreasing smoothing scales. However, this improvement starts to 
saturate at $\sim 10 \hMpc$ in matter field. At smaller smoothing,
the ES3 $r(k)$ starts to get more distorted
at intermediate $k \sim 0.1 \hMpc^{-1}$ scales.  

\begin{figure}
    \centering
    \includegraphics[width=0.98\columnwidth]{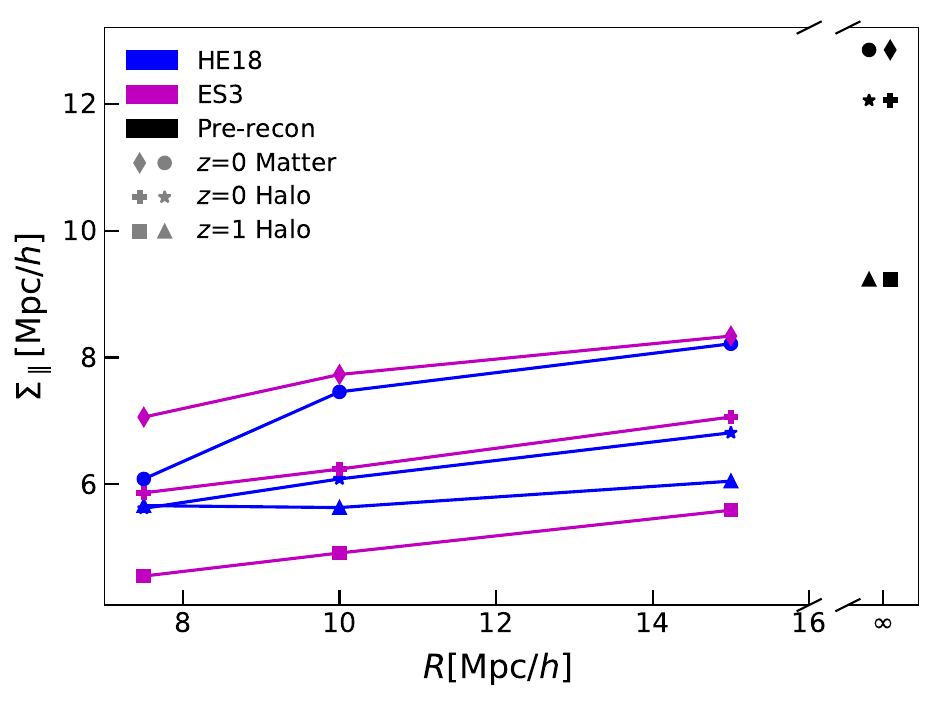}
    \vspace{0.11in}
    \caption{Estimated $\Sigma_{\parallel}$ as a function of smoothing scale,
    from fitting the monopole and quadrupole of the propagator with model 
    presented in the text. Points connected by blue lines are for the HE18 method,
    while those with magenta lines are for the ES3 method. From top to bottom, 
    the lines correspond to the $z=0$ matter field, the $z=0$ lower mass halo field, 
    and the $z=1$ lower mass halo field. 
    Also shown are the values of $\Sigma_{\parallel}$
    before reconstruction (at a smoothing scale of $\infty$). Reconstruction 
    clearly reduces the overall damping, although we do not find a clear trend 
    with the type of algorithm used. 
    }
    \label{fig:Sigpp_sm}
\end{figure}

Characterizing $G(k)$ is more challenging, since it is not bounded between $-1$ and $1$.
In particular, as can be seen from the functional form of the propagator presented in Eqn.~\ref{eqn:modified_gaussian},
the propagator is expected to exceed 1 due to RSD as one looks along the line of sight. 
Fig.~\ref{fig:halo12513_z0_zspace_gk_gkangle} shows $G(k)$ as a function of angle to the line of sight
for a fixed smoothing scale of 15 $\hMpc$ for the HE18 algorithm. The ``bump'' in the propagator 
due to RSD can clearly be seen. Furthermore, this bump gets more pronounced as one increases the 
smoothing scale, as expected from the Gaussian model for the propagator. 
Therefore, we fit the $l=0$ and $l=2$ multipoles of the propagator to our 
model (Eqns.~\ref{eqn:modified_gaussian}-~\ref{eqn:G_l}) and use the fitted parameters to characterize the impact of reconstruction. 
Fig.~\ref{fig:Sigpp_sm} shows these results. As with $r(k)$, we see a decrease in the damping 
parameters with decreasing smoothing scale (corresponding to a better cross-correlation), 
and that the improvements are most noticeable going from the case of the unreconstructed 
density field to any of the reconstructed cases. 
The differences between the two algorithms are less clear. 
We present the fits for $G(k)$ (with 7.5 $\hMpc$ smoothing scale) in Table~\ref{tab:prop_fits}. The ES3 model appears to perform better for modes perpendicular to the line of sight, but the trend is unclear in the parallel to the line of sight. 
However, the differences between the two algorithms are small, relative to the improvements 
over the unreconstructed field. 
Table~\ref{tab:prop_fits} summarizes the fitted parameters of the 
propagator for the different fields and algorithms. 

\begin{table}
\centering
\begin{tabular}{ccccc}
\hline
field & redshift & sample & $\Sigma_{\perp}$ & $\Sigma_{\parallel}$ \\ \hline
\multirow{6}{*}{matter} & \multirow{3}{*}{0} & pre-recon & 8.1 & 12.8  \\
 &  & HE18  & 4.5 & 6.1  \\
 &  & ES3  & 4.1 & 7.1  \\ \cline{2-5} 
 & \multirow{3}{*}{1} & pre-recon & 5.2 & 9.9  \\
 &  & HE18 & 3.1 & 5.6 \\
 &  & ES3 & 2.8 & 5.8    \\ \hline
\multirow{6}{*}{halo lower mass bin} & \multirow{3}{*}{0} & pre-recon & 7.8 & 12.1  \\
 &  & HE18 & 3.6 & 5.6 \\
 &  & ES3 & 3.1 & 5.9   \\ \cline{2-5} 
 & \multirow{3}{*}{1} & pre-recon & 4.8 & 9.2  \\
 &  & HE18 & 2.4 & 5.1  \\
 &  & ES3 & 1.9 & 4.6   \\ \hline
\multirow{6}{*}{halo higher mass bin} & \multirow{3}{*}{0} & pre-recon & 7.6 & 11.8   \\
 &  & HE18  & 3.7 & 6.2  \\
 &  & ES3 & 3.1 & 5.9    \\ \cline{2-5} 
 & \multirow{3}{*}{1} & pre-recon & 4.7 & 9.1 \\
 &  & HE18  & 2.7 & 5.7 \\
 &  & ES3 & 2.1 & 4.8   \\ \hline
\end{tabular}
\caption{The damping scales from the propagator for matter and halo fields in redshift space for both
reconstruction methods at $z=0$ and 1, using the modified Gaussian model described in the text.
The fits here are for the case of 7.5 $\hMpc$ smoothing. We find a mild dependence on the 
smoothing scale, summarized in Fig.~\ref{fig:Sigpp_sm}.
}
\label{tab:prop_fits}
\end{table}

\begin{figure*}
    \centering
    \includegraphics[width=0.99\columnwidth]{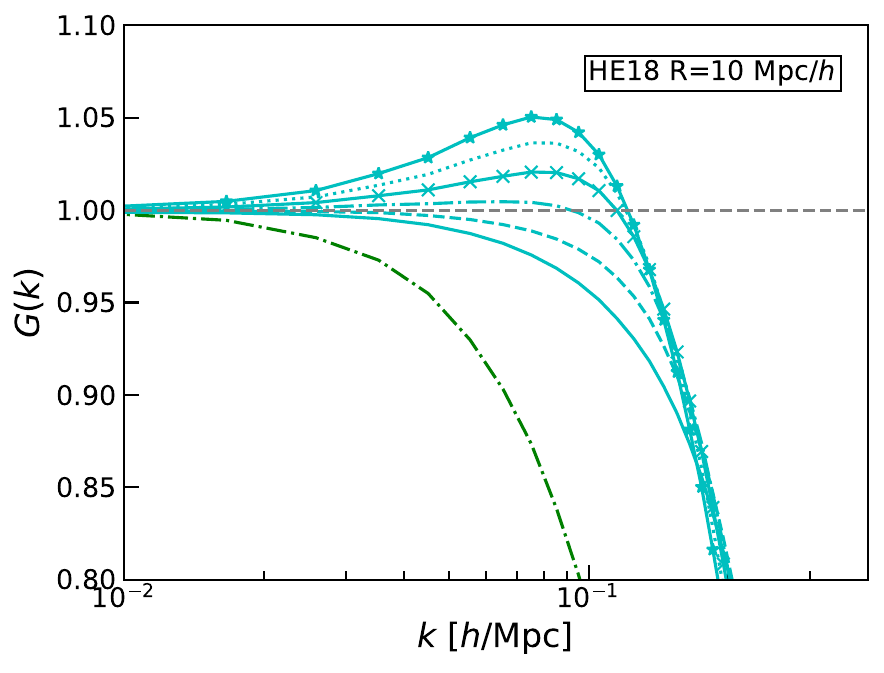}
    \includegraphics[width=1.01\columnwidth]{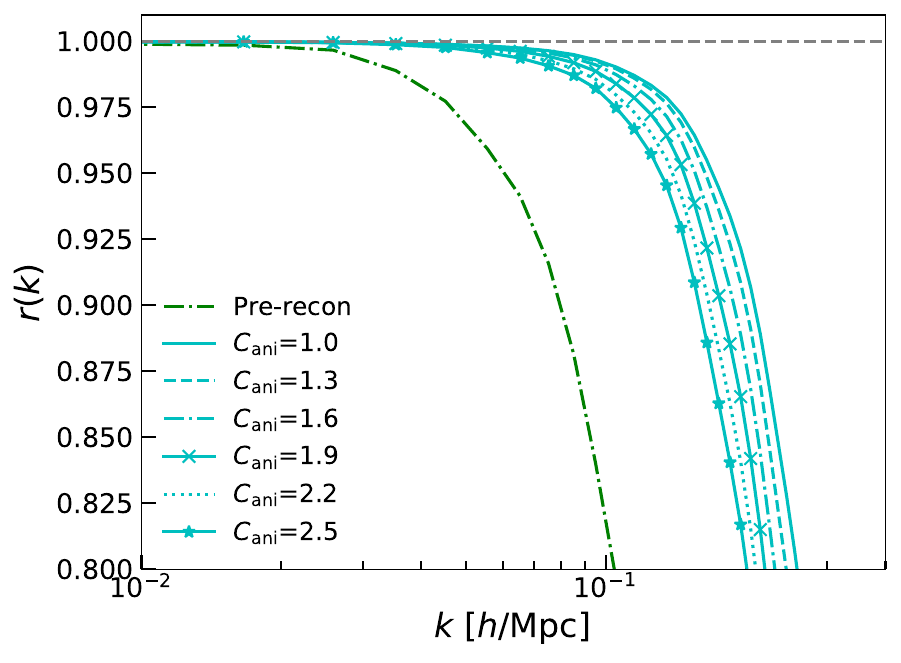}
    \caption{The propagator $G(k)$ and cross-correlation coefficient $r(k)$ of the matter field at $z=0$ in
    redshift space with anisotropic smoothing from HE18. All of these cyan lines
    have 10 $\hMpc$ smoothing perpendicular to the line-of-sight direction. The
    solid, dashed, dash-dotted, cross, dotted, and starred lines represent cases
    with $C_{\rm ani}$=1.0 (isotropic), 1.3, 1.6, 1.9, 2.2, and 2.5. 
    In general, increasing the anisotropy in the smoothing reduces $r(k)$, and 
    increases the RSD feature in $G(k)$ discussed earlier. For this particular case, 
    we find the best agreement with the initial conditions for $C_{\rm ani} \sim 1.6$.
    The trends in both $G(k)$ and
    $r(k)$ here also present in ES3.  }
    \label{fig:matter_z0_zspace_ani}
\end{figure*}

The angle dependence of the propagator suggests exploring the impact of an 
anisotropic smoothing kernel before reconstruction. Fig.~\ref{fig:matter_z0_zspace_ani}
shows $G(k)$ and $r(k)$ as a function of the anisotropic smoothing parameter $C_{\rm ani}$ 
defined above, with smoothing along perpendicular line of sight fixed at 10 $\hMpc$. We find that including a moderately larger smoothing scale along the line of sight 
can result in an improved $G(k)$ closer to 1, without a significant degradation in $r(k)$. 
The figure shows this effect for the HE18 algorithm and the matter density field, but 
similar results can be seen for other tracers and the ES3 algorithm. However, the 
exact degree of smoothing/anisotropy that is optimal does appear to depend on the tracer 
used. So, one way to optimize the smoothing for $G(k)$ can be finding a relatively small isotropic smoothing scale such that the bump does not present and then finding an $C_{\rm ani}$ that gets $G(k)$ closer to 1. Such an optimization is not possible with data, but must be done with simulations. One would need to test the robustness of the final results to these choices as well.

Finally, we note that both the HE18 and ES3 algorithms perform similarly when 
benchmarked against the unreconstructed density field. In general, the HE18 
algorithm performs somewhat better in the $r(k)$ statistic, and appears to 
be more stable when reducing the smoothing scale. This is likely a reflection of the 
iterative nature of the HE18 algorithm gradually reducing the smoothing scale.
Using too small a smoothing scale risks a breakdown in the simple model used to 
reconstruct the displacement field.

\subsection{Power spectrum}\label{sect:two-point}

\begin{figure*}
    \centering
    
    \includegraphics[width=0.99\columnwidth]{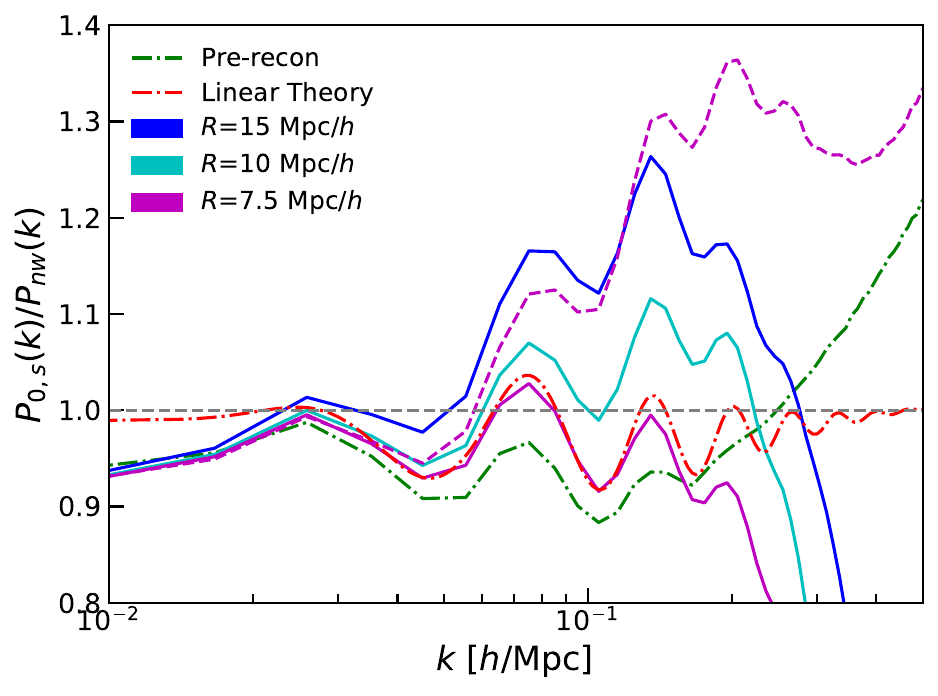}
    \includegraphics[width=1.01\columnwidth]{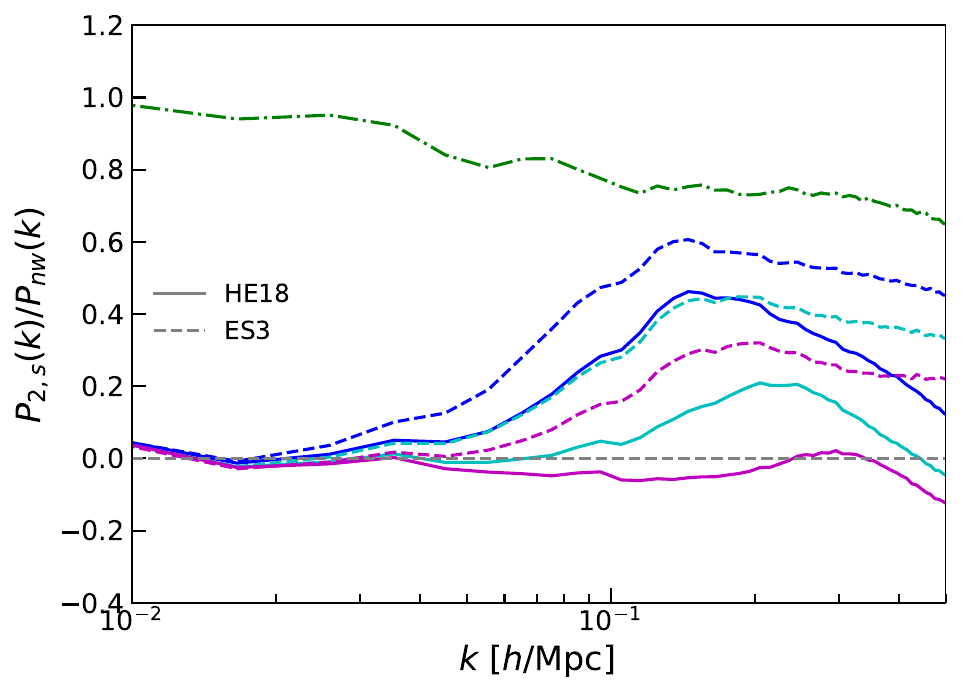}
    \caption{
        The monopole (left) and quadrupole (right) power spectra for the lower mass halo field at $z=0$ in redshift space. 
        To highlight the BAO feature, we divide the power spectra by a smooth power spectrum that approximately matches the shape of the 
        linear theory without any BAO features present. 
        When plotting the unreconstructed fields, we also account for linear redshift space distortions using the Kaiser approximation,
        explaining why the ratio of power spectra goes to 1 on large scales. Since our default implementation of the HE18 and ES3 
        models correct for RSD, we do not need anything further for the reconstructed fields.
        The solid lines show the HE18 algorithm, while
        the dashed lines are the ES3 algorithm. The green dash-dotted line shows the unreconstructed power spectra. 
        The left panel also shows the linear theory power spectrum (red dash-dotted lines). The different colors correspond to different smoothing scales:
        blue: 15 $\hMpc$, cyan: 10 $\hMpc$, magenta: 7.5
        $\hMpc$. 
        For clarity, we only show the 7.5 $\hMpc$ smoothing for the ES3 monopole. Larger smoothing scales 
        show even larger residual nonlinear effects. We observe that the HE18 algorithm 
        with a smoothing scale of 7.5 $\hMpc$ results in a monopole power spectrum that most closely matches linear 
        theory (out to $k\sim 0.1 \hMpc^{-1}$), and a quadrupole that is closest to zero (consistent with the removal of RSD).
        }
    \label{fig:halo12513_z0_zspace_gkrkp0p2}
\end{figure*}

Next we examine the monopole power spectrum of the reconstructed field. The
patterns in monopole across different fields are similar, so again we show the
lower mass bin of the halo fields at $z=0$ as an example. The left panel of
Figure~\ref{fig:halo12513_z0_zspace_gkrkp0p2} shows the monopole before/after
reconstruction with various smoothing scales in redshift space. 
While both the HE18 and ES3 algorithms effectively reconstruct the BAO features 
(that are more prominently seen compared with the unreconstructed field), the 
ES3 algorithm does not capture the broadband shape of the power spectrum as well. 
In particular, we see excess power at $k \sim 0.1 \hMpc^{-1}$ for the ES3 algorithm
and this excess power increases with increasing smoothing scale (not shown 
in the figure for clarity). By contrast, the HE18 algorithm does a better job of 
reconstructing the broadband shape of the power spectrum, especially with 
smaller smoothing scales. 
This is consistent with the results from the propagator 
and cross-correlations that we discussed earlier. 
The trend of power spectrum decreasing with decreasing smoothing scale continues with $R=5 \hMpc$. However, several cases with $R=5 \hMpc$ are not stable (iterations do not converge), so we do not include results of this smoothing scale in this paper.

We also examine the performance of the two algorithms in removing redshift space
distortions, as measured by the quadrupole power spectrum. In redshift space,
the quadrupole is non-zero due to redshift space distortions; on large scales, from 
the Kaiser effect and on small scales, from 
Finger-of-God effects. 
A perfect reconstruction would ideally remove all the
redshift space distortions due to Kaiser effect, returning a vanishing
quadrupole on large scales. The right panel of
Figure~\ref{fig:halo12513_z0_zspace_gkrkp0p2} shows the quadrupole power
for the lower mass bin halo field at $z=0$ in redshift space. We find HE18 performs
significantly better than ES3, bringing the quadrupole closer to zero. 
Similar behavior is seen for different tracers at different redshifts.
Removing redshift space
distortions is HE18's strongest advantage in our analyses. This improvement in removing redshift space distortions is also indicated in \citet[]{Duan19} with analysis in configuration space.

\section{Fitting BAO parameters}
\label{sec:BAO-fit}

The previous sections compared the two reconstruction methods at the 
level of the measured two-point statistics, especially in cross-correlation
with the initial density field. These comparisons are relevant if the 
goal is to accurately estimate the initial linear density field. However, 
the initial motivation behind reconstruction was to increase the S/N of 
BAO measurements. This is a weaker requirement than reconstructing the initial 
density field - we just require that we measure the BAO distance scale 
with a smaller bias and error than would be possible with the unreconstructed 
data (c.f. recent work aiming to recover the linear density field for a larger scale range, e.g. \citet[][]{cnn,Shallue23}). In what follows, we compare the inferred distance errors and biases 
for both reconstruction methods. 

\subsection{Method}

\subsubsection{Power Spectrum Template}

The traditional approach to measuring BAO distances is to 
compare the observed power spectrum/correlation function with a 
shifted/distorted theoretical template of the same. The shifts/distortions 
measure distances (as we review below), while 
the template captures the effects of nonlinear evolution, redshift 
space distortions and reconstruction.

While it is
possible to use a more theoretically motivated template for standard
reconstruction \citep[e.g.,][]{Padmanabhan09,Noh09,White14,Chen19,Seo21}, no such modelling
exists as yet for HE18 
(c.f. \citet[][]{Seo21} for the
treatment for a different iterative reconstruction method). 

For the uniformity of comparisons, we use a phenomenologically motivated 
template \citep{Eisenstein07b,Anderson12,Anderson14,Beutler17, Gil20} that 
has formed the basis of most recent BAO analyses. One of the results of this 
work is therefore the tuning and validation of this template for the HE18 
reconstruction method.

The 2D power spectrum
template is
\begin{equation}
    P^{\rm temp}(k,\mu)=(1+\beta\mu^2\Sigma_R)^2F(k,\mu,\Sigma_{s})P^{\rm dw}(k,\mu).
\end{equation}
The first term describes the Kaiser effect caused by galaxies infalling towards
the cluster center on large scales \citep{Kaiser87} and $\beta=f/b$, where $f$
is the growth factor and $b$ is linear galaxy bias. We use a
$\Sigma_R$-parameter to account for the removal of RSD in reconstruction:
\begin{equation}
\Sigma_R=1-\exp[-k^2R^2/2].
\end{equation}
$\Sigma_R$ is set to 1 before reconstruction \citep[][]{Beutler17,Seo16b}. 
This function captures the effect of the smoothing scale used in 
reconstruction. Note that the addition of $\Sigma_R$ is identical to 
what we used in the modified Gaussian model for the reconstructed propagator.
While $R$ could be simply fixed to the smoothing scale, we find that we get 
better fits when we allow it to vary. This is especially true for the 
HE18 algorithm, where the smoothing scale is not fixed, but gradually 
decreases over the course of the algorithm's iterations.
The second term represents the
finger-of-God (FoG) effect caused by random velocities within virialized
clusters on small scales \citep{Park94,Peacock94}. We use the model
\begin{equation}
F(k,\mu,\Sigma_s)=\frac{1}{(1+k^2\mu^2\Sigma_s^2/2)^2}, 
\end{equation}
where $\Sigma_s$ is the FoG parameter characterizing the dispersion due to
random peculiar velocities. The de-wiggled power spectrum $P^{\rm dw}$ has the form
\begin{equation}
\begin{split}
    P^{\rm dw}(k,\mu)&=P^{\rm nw}(k)\times\\
    & \left(1+(O^{\rm lin}(k)-1)\times\exp\left[-\frac{1}{2}\left(k^2\mu^2\Sigma_{\parallel}^2+k^2(1-\mu^2)\Sigma_{\perp}^2\right)\right] \right),
\end{split}
\end{equation}
where $O^{\rm lin}(k)$ is the ratio of the linear power spectrum and the linear
power spectrum with BAO wiggles smoothed out: $O^{\rm lin}(k)= P^{\rm
lin}(k)/P^{\rm nw}(k)$. Our $P^{\rm nw}(k)$ is a smoothed fit with polynomials
for the no-wiggle power spectrum form of \citet[][]{Eisenstein98}. The
exponential term on the far right models the damping of BAO due to nonlinear
evolution. In redshift space, this damping is anisotropic, and
$\Sigma_{\parallel}$ and $\Sigma_{\perp}$ are the streaming scales along and
perpendicular to line of sight, respectively. For real space, the redshift-space
associated terms all vanish, including $f$ (in the Kaiser term) and the
high-order multipoles,
and the streaming scales reduce to one parameter: $\Sigma_{\parallel}=\Sigma_{\perp}$.

We expand this 2D template into Legendre multipoles 
\begin{equation}
    P^{\rm temp}_{l}(k)=\frac{2l+1}{2}\int_{-1}^{1}P^{\rm temp}(k,\mu)L_{l}(\mu)\de \mu.
\end{equation}
where $L_{l}$ is the Legendre polynomial, with $l$ denoting the degree. We use
both monopole and quadrupole in the fitting procedure. 

As in standard BAO analyses, we also introduce a set of nuisance parameters to
account for errors in the fiducial cosmology, observational systematics, and
mismatches between the broadband shape of our template and the observations. Our
final functional form is
\begin{equation}
    P^{\rm fit}_{l}(k)=B^2P^{\rm temp}_{l}(k)+\mathcal{A}_{l}(k),
\end{equation}
where 
$B^2$ characterizes the linear galaxy bias. 
Note that we only fit one $B^2$ parameter for both monopole and quadrupole in
redshift space.  $\mathcal{A}_{l}$ are polynomials with nuisance parameters, and
we use 
\begin{equation}
\mathcal{A}_{l}(k)=A_{l,1}k^2+A_{l,2}k+
   A_{l,3}+\frac{A_{l,4}}{k}+\frac{A_{l,5}}{k^2}+\frac{A_{l,6}}{k^3} 
\end{equation}
where $A_{l,1}$-$A_{l,6}$ are nuisance parameters for each multipole power
spectrum. Including all these polynomial terms provides reasonable $\chi^2$ fits.

\subsubsection{Distance parameters}

To measure $D_A(z)$ and $H(z)$, one needs to assume a fiducial
cosmology to translate a measured sky position and redshift to Cartesian
coordinates. This fiducial cosmology may deviate from the true cosmology (in the
case of simulations, the cosmology used to generate the simulation data) one
aims to measure. We can measure this deviation by estimating two parameters,
isotropic dilation, $\alpha$, and anisotropic warping, $\epsilon$
\citep[][]{Padmanabhan08}. However, these two parameters couple the cosmological
parameters, $D_{A}(z)$ and $H(z)$, that we aim to measure. So we compute another
set of parameters, $\alpha_{\parallel}$ and $\alpha_{\perp}$, which decouple the
cosmological parameters and are thus favored in observational analysis recently
\citep[e.g.,][]{Beutler17}. In real space, we still measure $\alpha$ ($\epsilon$
vanishes). 
These two sets of parameters are related to $D_A(z)$ and $H(z)$ in the form of
\begin{equation}
    \alpha=\left[\frac{D_A^2(z)H_f(z)}{D_{A,f}^2(z)H(z)}\right]^{1/3}\frac{r_{s,f}}{r_s},
    1+\epsilon=\left[\frac{H_f(z)D_{A,f}(z)}{H(z)D_A(z)}\right]^{1/3},
\end{equation}
\begin{equation}
    \alpha_{\parallel}=\frac{H_{f}(z)r_{s,f}(z)}{H(z)r_s(z)},
    \alpha_{\perp}=\frac{D_A(z)r_{s,f}(z)}{D_{A,f}(z)r_s(z)},
\end{equation}
where $r_s$ is sound horizon and subscript $f$ stands for fiducial. These two
sets are related in this form: $\alpha_{\parallel}=\alpha(1+\epsilon)^2$ and
$\alpha_{\perp}=\alpha/(1+\epsilon)$. When the fiducial cosmology perfectly
matches the true cosmology, $\alpha=1$ and $\epsilon=0$, and
$\alpha_{\parallel}=\alpha_{\perp}=1$. Measuring these parameters in simulations
allows us to estimate any biases in the reconstruction, to forecast estimated
errors, and to compare the efficacy of different reconstruction schemes.

\subsubsection{Fitting Procedure}

With the above construction, we proceed to fitting by minimizing $\chi^2$,
following an approach similar to \citet{Xu12} and \citet[][]{Xu13}. 
We have in total 13 parameters in redshift
space and 7 parameters in real space.
We report
results without setting any prior. We have tested our fitting procedure with a
Gaussian prior for $B^2$ centered at our expected bias (1 for HE18's
post-reconstruction power spectrum, because HE18 removes bias) with 10 \%
error $\sigma_{B}$. Adding a bias prior does not give better $\chi^2$ fits. We
compute the $\chi^2$ goodness of fit as follows: $\chi^2=({\bf m} - {\bf
d})^{\rm T}C^{-1}({\bf m}-{\bf d})$, where {\bf m} is the model (fitted) vector
and {\bf d} is the data vector and $C$ is the covariance matrix.

We compute the posterior distribution for $\alpha_{\perp}$ and
$\alpha_{\parallel}$ on an $\alpha_{\perp}-\alpha_{\parallel}$ grid between
$0.9<\alpha_{\perp}<1.1$ and $0.85<\alpha_{\parallel}<1.15$, with step size
0.0025 for both $\alpha_{\perp}$ and $\alpha_{\parallel}$.
Since $B^2$ and $\mathcal{A}_{l}$ are linear parameters, we analytically
marginalize these out. The fits for $\alpha$ and $\epsilon$ are robust against
variations of $\Sigma_{\perp}$ and $f$ values. For example, \citet{Xu13} found
that a 10\% change in $\Sigma_{\rm nl}$ changes the maximum likelihood by less
than 1\% in real space. Therefore, we fix the pre-reconstruction
$\Sigma_{\perp}$ at 8 $\hMpc$ for $z=0$ and at 5 $\hMpc$ for $z=1$. We fix the
post-reconstruction $\Sigma_{\perp}$ at 4 $\hMpc$ for $z=0$ and at 3 $\hMpc$ for
$z=1$. We set $\Sigma_{\parallel}=(1+f)\Sigma_{\perp}$ for pre-reconstruction,
with $f$ corresponding to the value at each redshift, and
$\Sigma_{\perp}=\Sigma_{\parallel}$ after reconstruction. $f$ at the true value
is 0.5320 ($z=0$) or 0.8773 ($z=1$) in redshift space. With these parameters
set, we first fit two parameters, $\Sigma_R$ and $\Sigma_s$, at
($\alpha$,$\epsilon$)=(1,0) for each simulation pair. Then we input the unique
$\Sigma_R$ and $\Sigma_s$ values together with other parameters to perform
fitting on the $\alpha_{\perp}-\alpha_{\parallel}$ grid.

We calculate the covariance matrix for pre-reconstruction real space matter
fields directly from \Quijote's 15000 fiducial cosmology simulations. These
simulations have a lower resolution than the ones we use for reconstruction
($512^3$ CDM particles in a 1 $h^{-1}$Gpc box), but the cosmology is the same. For all other
cases, including post-reconstruction real space matter fields and all halo
fields, we calculate the analytic Gaussian covariance matrix \citep{Chudaykin19}
using our own calculated multipole power spectra, where bias and shot noise are
embedded in.

\subsection{Results}

Tables~\ref{tab:fits_realspace} and~\ref{tab:fits_zspace} summarize our fitting
of the BAO parameters in real and redshift space. For each case, we average two
simulations and fit 50 batches in 37 $k$ bins (37 $k$'s for each multipole)
between $k=0.03\hMpc$ and $k=0.4\hMpc$. Hence, we have estimates of ``scatter'',
the fluctuation of $\alpha$ mean among 50 batches, for our parameters. For real
space, we present {\it weighted} mean of $\alpha$, ``$\alpha$ scatter'', and an
estimate of the {\it median} of the ``$\alpha$ standard deviation'', which is
computed from the $\alpha$ grid described above. Similarly, redshift space
statistics are the weighted mean, scatter, median of standard deviation of both
$\alpha_{\parallel}$ and $\alpha_{\perp}$. We find that in most cases, the fits
are stable within a range of smoothing scales; in some cases, 15$\hMpc$ gives
large errors but 10$\hMpc$ and 7.5$\hMpc$ provide comparable results. Since the
BAO parameter estimates are robust against different smoothing scales, we
present results using HE18 with 7.5 $\hMpc$ effective smoothing and ES3 also
with 7.5 $\hMpc$ smoothing throughout, except for specific analysis we will
discuss later in Section~\ref{sec:discussion}. In Table~\ref{tab:fits_realspace}, the scatter and standard
deviation are generally consistent. In real space, HE18 and ES3 give comparable
fitting results, and both improve the error in $\alpha$ from before
reconstruction by a factor of two or three in matter field, and about 60\% in
halo fields. 

\begin{figure}
    \centering
    \includegraphics[width=\columnwidth]{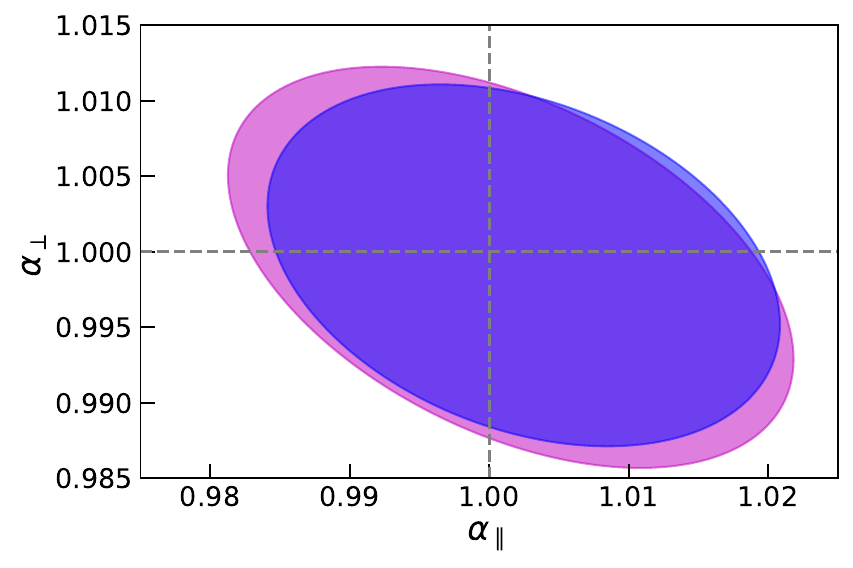}
    \includegraphics[width=\columnwidth]{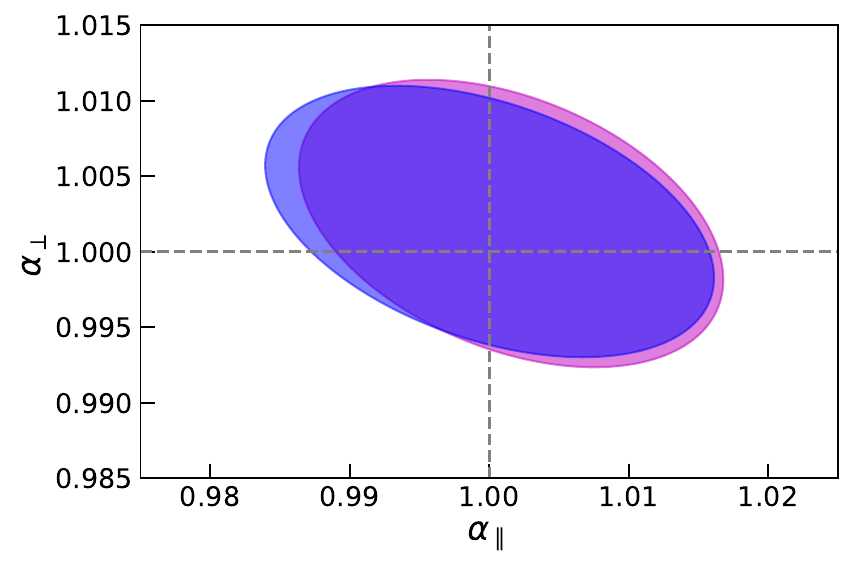}
    \caption{$\alpha_{\parallel}$ and $\alpha_{\perp}$ 1-$\sigma$ confidence interval for redshift space lower mass bin halo field at $z=0$ (top) and $z=1$ (bottom). Blue: HE18, magenta: ES3. We use scatter as the error and the values are present in Table~\ref{tab:fits_zspace}. We calculate the covariance between the two parameters from 50 fits. The two contours show that HE18 and ES3 are consistent and the fitting results are also consistent with the fiducial cosmology. }
    \label{fig:contour}
\end{figure}

In redshift space, the performance of HE18 and ES3 are again consistent. Both
algorithms improve the error estimates by about a factor of 2-4. The improvement
is slightly more significant at $z=0$ than at $z=1$. To show the size of the
errors of the BAO parameter estimates after reconstruction, we present the
confidence interval contours using the scatters of the estimates shown in
Figure~\ref{fig:contour}, for the lower mass halo field as an example. As shown
in this figure, the estimated $\alpha_{\parallel}$ and $\alpha_{\perp}$ are
consistent with the true values. In Table~\ref{tab:fits_zspace}, we also present
fitting results for another halo sample. We fix the halo number density at
$3\times10^{-4}$ $\Mpc^3$ at $z=0$ and 1; this way we can directly compare the
results at the two redshift. The results again show slightly better performance
at $z=0$. Both algorithms improve the error estimate of $\alpha_{\parallel}$ by
about a factor of 2.2 at $z=0$ and 1, but improve by a factor of 3 for
$\alpha_{\perp}$ at $z=0$ versus 2.2 at $z=1$, although the absolute errors are
more often larger at $z=0$.

We also examine the cross-correlation between the measured $\alpha$ after
reconstruction and the $\alpha$ of initial condition, using the 50 pairs of
simulations, in Table~\ref{tab:alpha_cross_summary}. The smoothing scale
presented is again 7.5 $\hMpc$. We find that HE18 are consistently more
correlated with initial condition in both real and redshift space and across
different fields. Moreover, often the smaller the smoothing scale, the better
the cross-correlation, shown in Table~\ref{tab:alpha_cross_smoothing} with real
space matter field at $z=0$ as an example. The observation that an $\alpha$
cross-correlation for a higher smoothing scale for HE18 is matched with a
smaller smoothing scale by ES3, if it can be matched, echos the trend in the
propagator results. Within one reconstruction
algorithm, the smaller the smoothing, the better the $\alpha$ cross-correlation
suggests that larger smoothing could add additional noise to the distance measures.
Recent studies \citep[e.g.][]{Chen19} argue a larger smoothing for the ES3 algorithm, for a better match with theory. 
However, we do not detect any bias in our BAO fits using a relatively small smoothing. For HE18, since the smoothing starts at 
a large value ($R_{\rm ini}=20\ \hMpc$) and gradually reduces to its final value, 
it may be possible to push to smaller smoothing scales. Hybrid algorithms 
that pair reconstruction algorithms like HE18 and ES3 with machine learning 
\citep{Shallue23, cnn} may also make the exact choice of smoothing scale less 
relevant. We leave a detailed study of these for later work.

\begin{table*}
\centering
\begin{tabular}{ccccccc}
\hline
field & redshift & sample & $\alpha$ mean & $\alpha$ scatter & $\alpha$ std  \\ \hline
\multirow{6}{*}{matter} & \multirow{3}{*}{0} & pre-recon & 1.0059 & 0.0138 & 0.0130  \\
 &  & {HE18 (7.5)} & {1.0007} & {0.0038} & {0.0047}  \\
 &  & {ES3 (7.5)} & {1.0010} & {0.0044} & {0.0054}  \\ \cline{2-6} 
 & \multirow{3}{*}{1} & pre-recon &  1.0031 & 0.0072  & 0.0077  \\
 &  & {HE18 (7.5)} & {1.0006} & {0.0037} & {0.0044} \\
 &  & {ES3 (7.5)} & {1.0007} & {0.0039} & {0.0046}  \\ \hline
\multirow{6}{*}{halo lower mass bin} & \multirow{3}{*}{0} & pre-recon & 1.0027 & 0.0152 & 0.0151  \\
 &  & {HE18 (7.5)} & {0.9997} &{0.0085} & {0.0080}  \\
 &  & {ES3(7.5)} & {0.9997} & {0.0088} & {0.0086}  \\ \cline{2-6} 
 & \multirow{3}{*}{1} & pre-recon & 1.0041  & 0.0101 & 0.0105 \\
 &  & {HE18 (7.5)} & {1.0013} & {0.0064} & {0.0067}  \\
 &  & {ES3 (7.5)} & {1.0016} & {0.0065} & {0.0070} \\ \hline
\multirow{6}{*}{halo higher mass bin} & \multirow{3}{*}{0} & pre-recon & 1.0107 & 0.0164 & 0.0168 \\
 &  & {HE18 (7.5)} &{1.0034} & {0.0110} & {0.0095} \\
 &  & {ES3 (7.5)} & {1.0037} & {0.0115} & {0.0105} \\ \cline{2-6} 
 & \multirow{3}{*}{1} & pre-recon & 1.0053 & 0.0127 & 0.0125 \\
 &  & {HE18 (7.5)} & {1.0003} & {0.0074} & {0.0085}  \\
 &  & {ES3 (7.5)} & {1.0006} & {0.0070} & {0.0088} \\ \hline
\end{tabular}
\caption{BAO parameter fits for matter and halo fields in real space for both
reconstruction methods at $z=0$ and 1. The after reconstruction entries
represent $\alpha$ constraints using smoothing scales 7.5 $\hMpc$ for both HE18
and ES3.}
\label{tab:fits_realspace}
\end{table*}

\begin{table*}
\centering
\begin{tabular}{ccccccccc}
\hline
field & \multicolumn{1}{l}{redshift} & sample & $\alpha_{\parallel}$ mean & $\alpha_{\parallel}$ scatter & $\alpha_{\parallel}$ std & $\alpha_{\perp}$ mean & $\alpha_{\perp}$ scatter & $\alpha_{\perp}$ std \\ \hline
\multicolumn{1}{c}{\multirow{6}{*}{matter}} & \multirow{3}{*}{0} & pre-recon & 1.0247 & 0.0423 & 0.0457 & 1.0019 & 0.0250 & 0.0211 \\
\multicolumn{1}{c}{} &  & HE18 (7.5) & 1.0028 & 0.0130 & 0.0121 & 1.0005 & 0.0055 & 0.0071 \\
\multicolumn{1}{c}{} &  & ES3 (7.5) & 1.0002 & 0.0142  & 0.0128 & 1.0009 & 0.0068 & 0.0080  \\ \cline{2-9} 
\multicolumn{1}{c}{} & \multirow{3}{*}{1} & pre-recon & 1.0173 & 0.0306 & 0.0326 & 0.9996 & 0.0149 & 0.0131 \\
\multicolumn{1}{c}{} &  & HE18 (7.5) & 1.0013 & 0.0119 & 0.0106 & 1.0005 & 0.0049 & 0.0066 \\
\multicolumn{1}{c}{} &  & ES3 (7.5) & 1.0015 & 0.0122 & 0.0117 & 1.0008 & 0.0054 & 0.0069 \\ \hline
\multirow{6}{*}{halo lower mass bin} & \multirow{3}{*}{0} & pre-recon & 1.0281 & 0.0461 & 0.0485 & 0.9967 & 0.0293 & 0.0248 \\
 &  & HE18 (7.5) & 1.0025 & 0.0185 & 0.0183 & 0.9991 & 0.0118 & 0.0121  \\
 &  & ES3 (7.5) & 1.0012 & 0.0199 & 0.0191 & 0.9992 & 0.0129 & 0.0126 \\ \cline{2-9} 
 & \multirow{3}{*}{1} & pre-recon & 1.0181 & 0.0325 & 0.0376 & 1.0018 & 0.0197 & 0.0164  \\
 &  & HE18 (7.5) & 1.0003 & 0.0159 & 0.0157 & 1.0018 & 0.0089 & 0.0100 \\
 &  & ES3 (7.5) & 1.0017 & 0.0153 & 0.0164 & 1.0020 & 0.0093 & 0.0105 \\ \hline
\multirow{6}{*}{halo higher mass bin} & \multirow{3}{*}{0} & pre-recon & 1.0199 & 0.0428 & 0.0549 & 1.0112 & 0.0279 & 0.0283 \\
 &  & HE18 (7.5) & 1.0064 & 0.0231 & 0.0223 & 1.0009 & 0.0144 & 0.0137 \\
 &  & ES3 (7.5) & 1.0069 & 0.0259 & 0.0230 & 1.0011 & 0.0148 & 0.0150 \\ \cline{2-9} 
 & \multirow{3}{*}{1} & pre-recon & 1.0256 & 0.0414 & 0.0435 & 0.9995 & 0.0228 & 0.0206 \\
 &  & HE18 (7.5) & 1.0083 & 0.0235 & 0.0226 & 0.9978 & 0.0107 & 0.0128 \\
 &  & ES3 (7.5) & 1.0073  & 0.0223 & 0.0203 & 0.9983 & 0.0103 & 0.0133 \\ 
 \hline
\multirow{6}{*}{halo fixed number density} & \multirow{3}{*}{0} & pre-recon & 1.0254 & 0.0429 & 0.0516 & 1.0035 & 0.0282 & 0.0246 \\
 &  & HE18 (7.5) & 1.0047 & 0.0198 & 0.0164 & 0.9998 & 0.0092 & 0.0107 \\
 &  & ES3 (7.5) & 1.0028 & 0.0199 & 0.0185 & 1.0000 & 0.0100 & 0.0116 \\ \cline{2-9} 
 & \multirow{3}{*}{1} & pre-recon & 1.0244 & 0.0398 & 0.0413 & 1.0002 & 0.0209 & 0.0182 \\
 &  & HE18 (7.5) & 1.0062 & 0.0190 & 0.0191 & 1.0000 & 0.0094 & 0.0107 \\
 &  & ES3 (7.5) & 1.0049 & 0.0173 & 0.0183 & 1.0005 & 0.0092 & 0.0108 \\
 \hline
\end{tabular}
\caption{BAO parameter fits in redshift space for matter and halo fields at both
redshift. After reconstruction values represent HE18 and ES3 both with 7.5
$\hMpc$ smoothing. }
\label{tab:fits_zspace}
\end{table*}

\begin{table*}
\centering
\begin{tabular}{cccccc}
\hline
field & space & redshift & pre-recon & HE18 & ES3 \\ \hline
\multirow{4}{*}{matter} & \multirow{2}{*}{real} & 0 & 0.3204 & 0.8595 & 0.8022 \\
 &  & 1 & 0.5555 & 0.9012 & 0.8999 \\ \cline{2-6} 
 & \multirow{2}{*}{redshift} & 0 & 0.3389 & 0.7738 & 0.6465 \\
 &  & 1 & 0.5008 & 0.8799 & 0.8324 \\ \hline
\multirow{4}{*}{halo lower mass bin} & \multirow{2}{*}{real} & 0 & 0.3142 & 0.5009 & 0.4462 \\
 &  & 1 & 0.4885 & 0.6865 & 0.6719 \\ \cline{2-6} 
 & \multirow{2}{*}{redshift} & 0 & 0.3151 & 0.4616 & 0.3728 \\
 &  & 1 & 0.4508 & 0.6782 & 0.6559 \\ \hline
\multirow{4}{*}{halo higher mass bin} & \multirow{2}{*}{real} & 0 & 0.4643 & 0.6121 & 0.6403 \\
 &  & 1 & 0.2774 & 0.4603 & 0.4219 \\ \cline{2-6} 
 & \multirow{2}{*}{redshift} & 0 & 0.4586 & 0.6358 & 0.6079 \\
 &  & 1 & 0.3282 & 0.5139 & 0.4826 \\ \hline
 \multirow{2}{*}{halo fixed number density} & \multirow{2}{*}{redshift} & 0 & 0.2104 & 0.4142 & 0.4307 \\
 & & 1 & 0.3752 & 0.5526 & 0.5979 \\ \hline
\end{tabular}
\caption{Cross-correlation of $\alpha$ post-reconstruction with the initial condition.}
\label{tab:alpha_cross_summary}
\end{table*}

\begin{table}
\centering
\begin{tabular}{lccc}
\hline
\backslashbox{method}{smoothing\\ ($\hMpc$)}
 & 15 & 10 & 7.5 \\ \hline
HE18 & 0.5776 & 0.7288 & 0.7738 \\
ES3 & 0.5016 & 0.6252 & 0.6465\\
\hline
\end{tabular}
\caption{Cross-correlation between $\alpha$ after reconstruction and that from initial condition for matter field in redshift space at $z=0$.}
\label{tab:alpha_cross_smoothing}
\end{table}

\section{Discussion}\label{sec:discussion}

\subsection{HE18 with ${\bf S}_l^{(2)}$ turned off}

One of the major differences between HE18 and the ES3 method is whether the
second order solution to the Euler-Poisson equation is used in the
reconstruction procedure. When the second order correction is turned off in
HE18, the theoretical assumptions are essentially the same for the two methods.
Therefore, we can isolate this theoretical input and evaluate the contributions
from other features in HE18, such as the iterative mechanism and annealing
smoothing.

\begin{figure*}
    \centering
    \includegraphics[width=0.99\columnwidth]{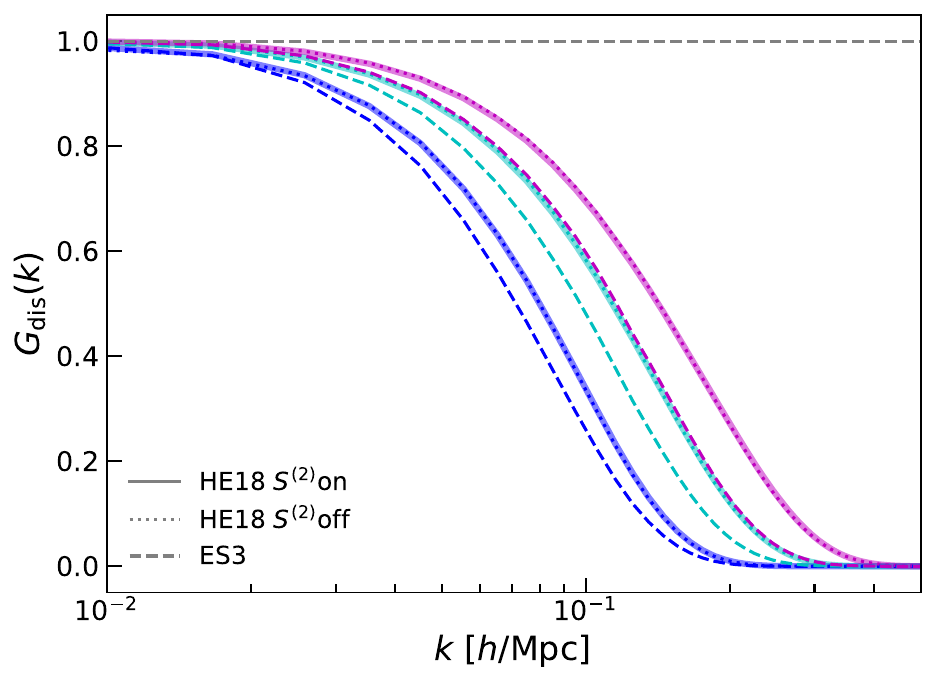}
    \includegraphics[width=1.01\columnwidth]{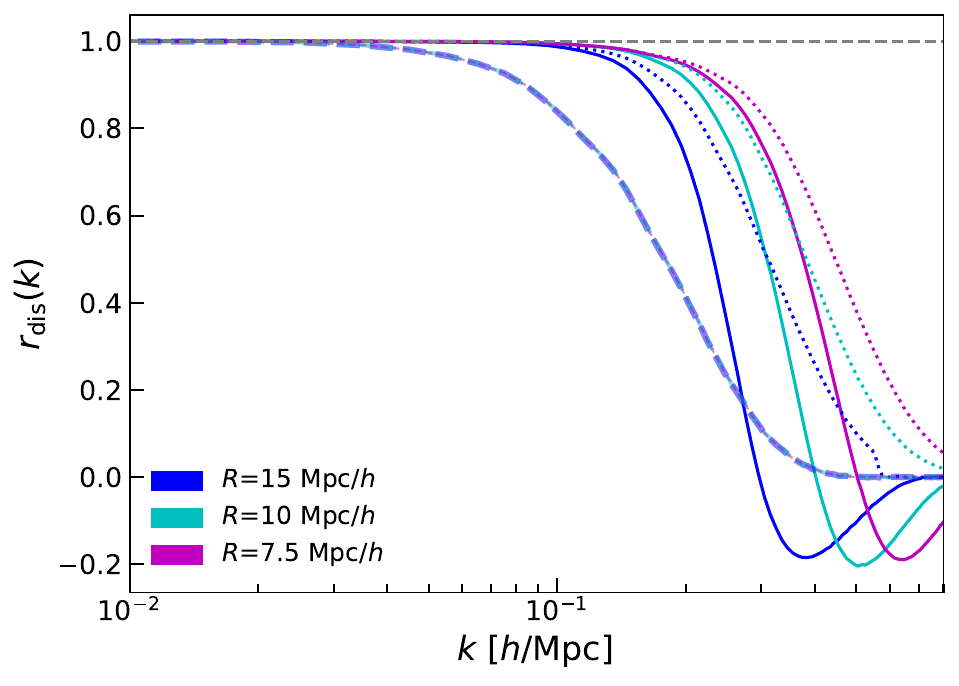}
    \caption{Displacement field cross-correlations by HE18 with $S^{(2)}$ turned
    on (solid) and off (dotted) and by ES3 (dashed) for the matter field in redshift space $z=0$. Some line widths are increased to show overlapping results. Both
    $S^{(2)}$ turned on and off for HE18 gives better results compared with ES3
    for a given smoothing scale. However, results of a smaller smoothing by ES3
    can be matched with a larger smoothing by HE18 in displacement propagator
    $G_{\rm dis}(k)$. In displacement propagator, the differences between
    $S^{(2)}$ turned on and off are negligible, but in displacement
    cross-correlation coefficient $r_{\rm dis}(k)$, the differences are more
    significant. Also, HE18 with $S^{(2)}$ turned off produces a better
    displacement cross-correlation. Different smoothing scales for ES3 does not
    make a difference in $r_{\rm dis}(k)$ because the smoothing kernel cancels
    out. Results of real space of matter field are similar.}
    \label{fig:dis_cross_s2off}
\end{figure*}

Since the second-order perturbation equations are used to determine the displacements, 
we start by comparing the observed displacement to the actual particle displacements
(the difference between final and initial positions).
Analogous to the density field, we construct the propagator and cross-correlation coefficient 
of the displacement fields by 
\begin{equation}
G_{\rm dis}(k)=\frac{\left<{\bf \tilde{S}}_l(k)\cdot {\bf \tilde{S}_{\rm
true}^{*}(k)} \right>}{\left<|{\bf \tilde{S}_{\rm true}(k)}|^2\right>}
\end{equation}
and 
\begin{equation}
r_{\rm dis}(k)=\frac{\left<{\bf
\tilde{S}}_l(k)\cdot {\bf \tilde{S}_{\rm true}^{*}(k)}
\right>}{\sqrt{\left<|{\bf \tilde{S}}_{l}(k)|^2 \right>\left<|{\bf
\tilde{S}_{\rm true}(k)}|^2\right>}}
\end{equation}
Figure~\ref{fig:dis_cross_s2off} shows that the propagator in redshift space
matter fields between ${\bf S}_{l}^{(s)}$ and ${\bf S}_{\rm true}$ and that
between ${\bf S}_{l}^{(1)}$ and ${\bf S}_{\rm true}$ are 
virtually indistinguishable, while for $r_{\rm dis}(k)$, adding in the second-order solution
appears to decorrelate the displacement field. 
By comparison, at the same smoothing scale, the ES3 method produces a displacement
field that is noticeably worse than either variant of HE18. The 
discrepancy in $G_{\rm dis}(k)$ may be mitigated by working at a smaller smoothing scale, 
but this is not the case for $r_{\rm dis}(k)$, which is independent of the smoothing.
\footnote{For the ES3 method, the displacement field is just a linear function 
of the density field in Fourier space, and so, this cancels out in the 
cross-correlation. This cancellation does not occur for HE18 due to the 
iterative nature of the algorithm.}
We believe that this improvement comes from the fact that the HE18 algorithm 
estimates the density field in Lagrangian space (iteratively), and uses this 
density to estimate the displacement field. Since the input to our 
perturbation theory equations is the density field in Lagrangian space, this
appears to be a more accurate approach.

Given the relatively small changes in the displacement field, it is not 
surprising that including/excluding the second-order terms make little 
difference in the final reconstructed field (in any of the metrics we 
have considered above). We therefore conclude that one can safely drop this 
correction in the algorithm with no degradation to the reconstruction. We also note that some previous studies that looked at second-order correction for reconstruction also find negligible differences \citep[e.g.][]{Seo10,Schmittfull17}.

One question this result raises is why second order displacement does not improve reconstruction of the initial condition, when higher-order displacement benefits reconstruction of the nonlinear field in forward modeling \citep[e.g.][]{Schmidt21}. We believe that this is intrinsic to the differences between forward and reverse modeling. Conceptually, the dominant displacement is the Zel’dovich term which translates the Lagrangian region to its final Eulerian position. This move is approximately invertible; one can move back to the initial position. The higher order terms then are responsible for the deformation of the Lagrangian region into its final Eulerian shape. However, it is not possible to disentangle this just from a measurement of the final density field. While the above is meant to be conceptual, we believe it elucidates why higher order perturbations might improve the final density field in forward modeling, but seem to have little effect here.

\subsection{Annealing anisotropic smoothing in redshift space}

\begin{figure*}
    \centering
    \includegraphics[width=\columnwidth]{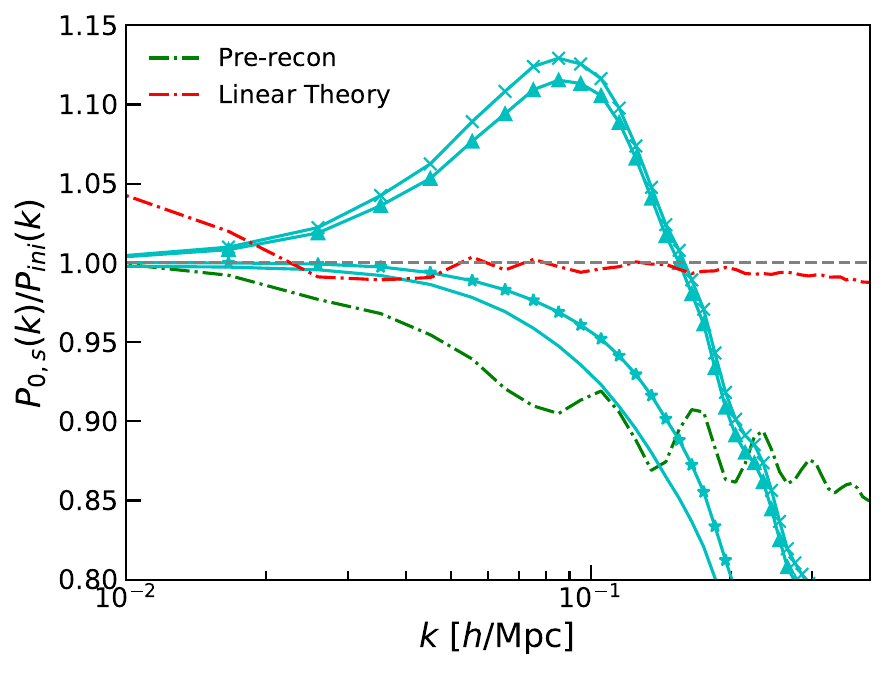}
    \includegraphics[width=\columnwidth]{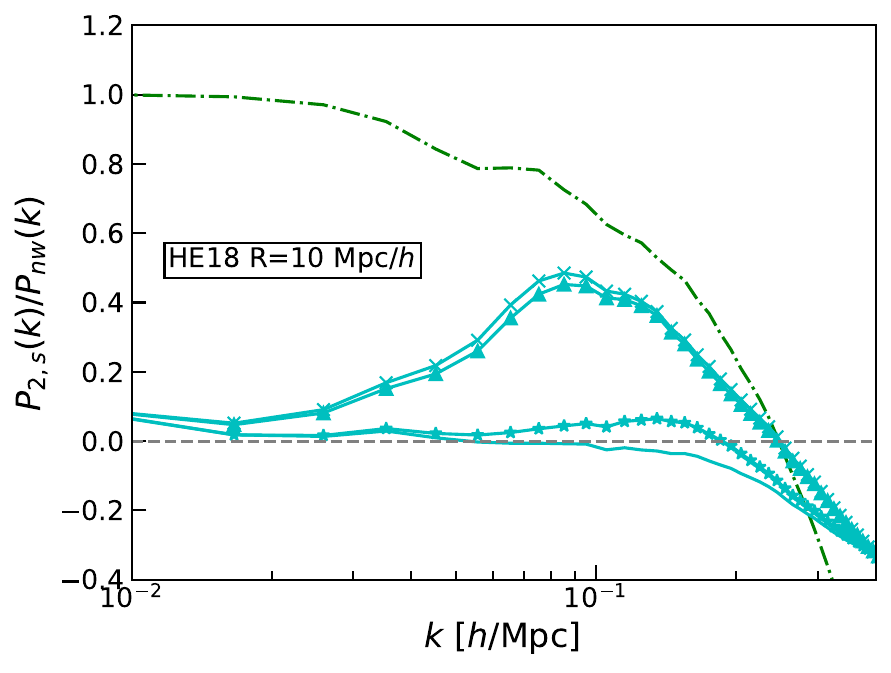}
    \includegraphics[width=0.99\columnwidth]{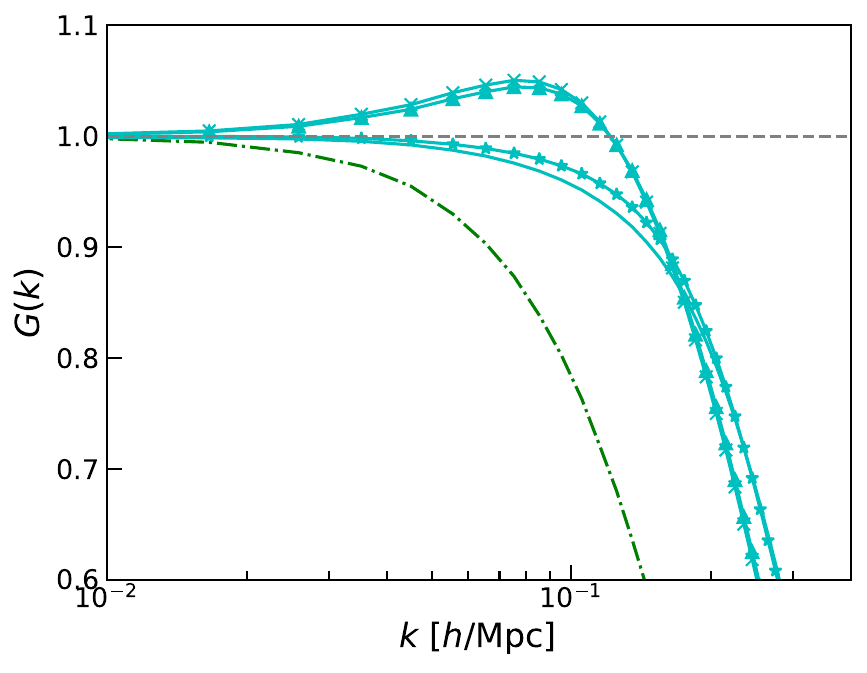}
    \includegraphics[width=1.01\columnwidth]{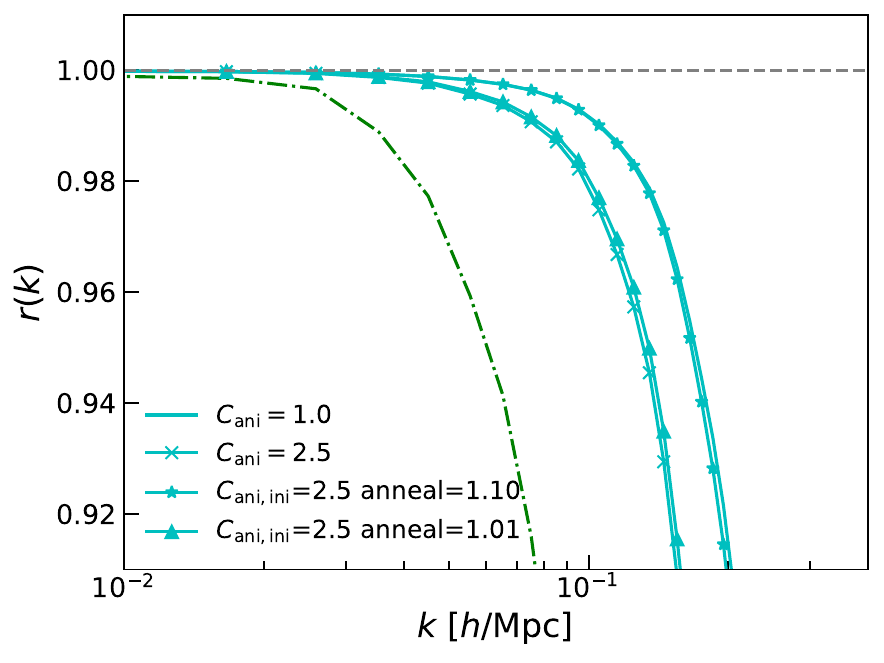}
    \caption{Monopole (upper left), quadrupole (upper right) power spectra, propagator (lower left) and cross-correlation coefficient (lower right) of annealing anisotropic smoothing for redshift space matter field using HE18 with 10 $\hMpc$ smoothing at $z=0$. Note that we divide the monopole power spectrum by the initial power spectrum, instead of no-wiggle power spectrum, to better show the pattern. Solid line is isotropic smoothing, crossed line is anisotropic smoothing with $C_{\rm ani}$=2.5 without annealing, and triangle and star lines are anisotropic smoothing with $C_{\rm ani,ini}$=2.5 with annealing rate equal to 1.01 and 1.1, respectively. Green dash-dotted line is before reconstruction. Red dash-dotted line is linear theory. In all but the cross-correlation coefficient,
    anisotropic smoothing with $C_{\rm ani,ini}$=2.5 and with a higher annealing rate (star line) gives the best results. }
    \label{fig:anneal_ani}
\end{figure*}

HE18 implemented 
the idea of using an anisotropic smoothing kernel in 
reconstruction. However, they kept the degree of anisotropy fixed 
over the course of the reconstruction. Since the intermediate steps 
in the reconstruction yield a density field with redshift space distortions 
reduced, it is interesting to ask whether reducing the degree of anisotropy 
with each iteration could improve the reconstruction. This is 
analogous to the annealing of the overall isotropic smoothing scale that 
works from the largest scales down to smaller scales. In order to implement 
this, we use a simple annealing schedule described in Sec.~\ref{sec:anisotropic_smoothing}.

Figure~\ref{fig:anneal_ani} shows results using annealing rates of 1.1 and 1.01
with a starting anisotropic factor 
$C_{\rm ani,ini}$=2.5 and 10 $\hMpc$ smoothing perpendicular to line of sight with redshift space matter
field at $z=0$. In both monopole and quadrupole power spectra and propagator,
$C_{\rm ani}$=2.5 with an annealing rate of 1.1 offers better performance 
as measured in the propagator and power spectra, and a similar cross-correlation
coefficient. In particular, the reduction in the anisotropic smoothing 
appears to be able to mitigate the RSD bump seen in the propagator. 
Using this annealing rate does not also degrade the errors on the BAO distance 
parameters. As with our previous results, this suggests that these variants 
of the algorithm might be unnecessary for BAO reconstruction, but could be 
useful for other applications.

\subsection{Anisotropic reconstruction in redshift space}\label{sec:RecAni}

\begin{figure*}
    \centering
    \includegraphics[width=\columnwidth]{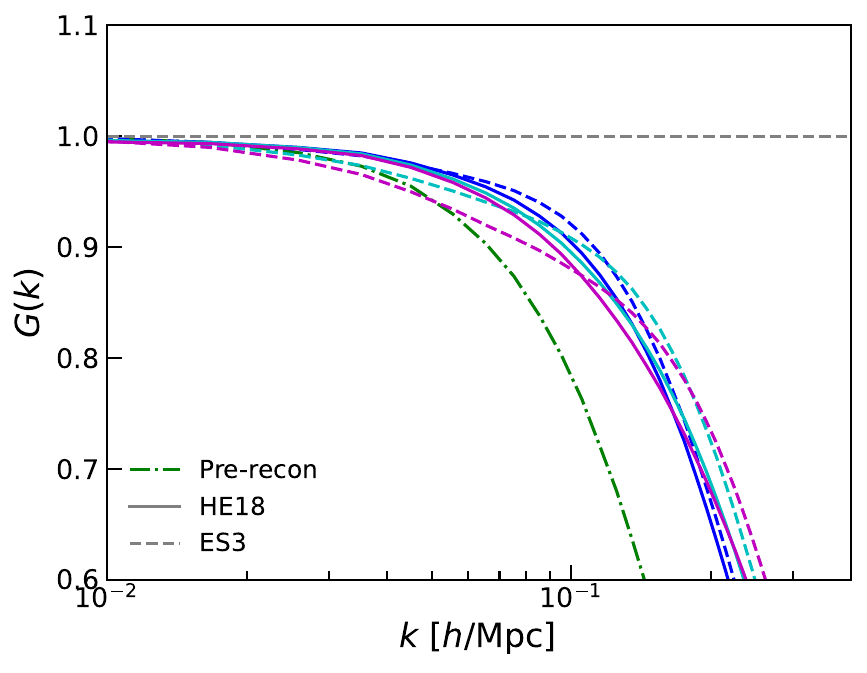}
    \includegraphics[width=\columnwidth]{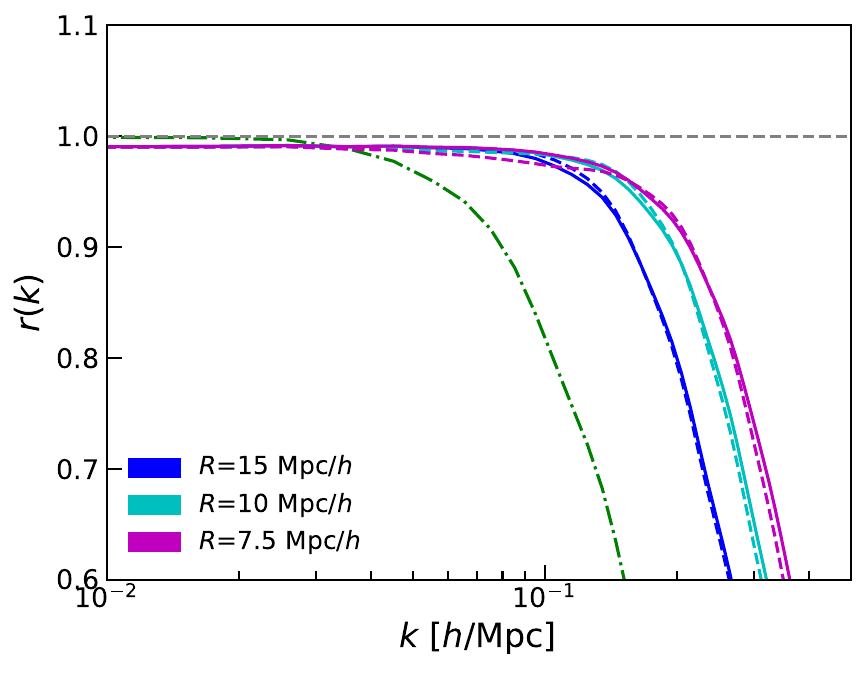}
   
    \caption{$G(k)$ and $r(k)$ of anisotropic reconstruction for matter field at
    $z=0$. Line styles and color scheme follow those in    Figure~\ref{fig:halo12513_z0_zspace_matter_z0_zspace_rk}.
    Different smoothing scales do
    not produce clear differences in the propagator, but do so in $r(k)$. The
    larger the smoothing scale, the faster it falls.}
    \label{fig:RecAni_gkrk}
\end{figure*}

We examine the anisotropic reconstruction using matter field at $z=0$, 
leaving the RSD signal in the data.
Figure~\ref{fig:RecAni_gkrk} shows $G(k)$ and $r(k)$ with both algorithms and
various smoothing scales. The RSD bump in the propagator is no longer present 
in these results, and we detect no clear trend with the smoothing scale, except that at the turnover, larger smoothing maintains slightly higher $G(k)$, 
although without a bump. 
This is roughly consistent with what we found in the isotropic case. 
The cross-correlation coefficient $r(k)$, however, does show a clear
improvement with a decreasing smoothing scale, with very little difference 
between HE18 and ES3.

We also perform BAO fits with these reconstruction results. We use the same
$\Sigma_{\perp}$ values as those used in isotropic reconstruction fits, but we
set $\Sigma_{\parallel}=(1+f)\sigma_{\parallel}$, where $f$ is not reduced,
after reconstruction. We also keep $R$-term at infinity. 
Comparing to isotropic
reconstruction, the fits for $\alpha_{\parallel}$ are comparable to those
isotropic counterparts. Both scatter and standard deviation of
$\alpha_{\parallel}$ by HE18 are similar to those by ES3. Fits for
$\alpha_{\perp}$ are also consistent with those in the corresponding isotropic
cases. 
Also, the improvement after reconstruction is still larger at $z=0$ than
at $z=1$, about a factor of 3.6 versus 2.6. 

The advantages of leaving the RSD signal in the reconstructed data appear 
to be to reduce various artifacts from the smoothing scale, allowing 
better modeling of these results in the case of the ES3 algorithm. 
We currently 
do not have a full model for the HE18 reconstructed field and so it is not
clear that the same modeling advantages transfer to this case, although 
e.g. the mitigation of the feature in the propagator is suggestive.
However, since many of these are broadband features, they do not significantly impact the BAO fits.

\subsection{HE18 vs ES3}
\label{sec:he18_vs_es3}

Our primary conclusion from this study is that the HE18 and ES3 methods are 
generally quite comparable in terms of performance across metrics that 
focus both on the density field as well as fitting the BAO feature. 
We find that the ES3 method can return similar results in the cross-correlation 
coefficients and propagators to HE18, albeit using a smaller effective smoothing scale. 
However, the HE18 method is more robust to the choice of smoothing scale, 
and so, has the potential to push down to smaller scales. The HE18 method 
does better at estimating the displacement field, and so, can better remove 
the RSD signal in the data. It also appears to do better at reconstructing the 
full broadband shape of the power spectrum, making it more useful for general 
LSS studies.

The two reconstruction methods show comparable results in BAO distance fitting
in both real and redshift space. If we correlate the $\alpha$ values post-reconstruction
with those measured from the initial density field, we find that HE18 does have a 
higher cross-correlation coefficient than ES3. This is consistent with the 
field-level $r(k)$ results above.

We find that HE18 is robust to the various parameter choices, and 
tracer populations. There are certain aspects where HE18 performs better than ES3; in
particular, HE18 is significantly better at estimating the large scale velocity field 
and removing RSD. Even though HE18 does
not present a substantial improvement in BAO fitting, it is a promising
method to be applied in other areas of large-scale structure analysis. 

\section{Conclusions}\label{sec:concl}


We closely study the new iterative reconstruction algorithm by \citet[][]{Hada18} in comparison to the standard algorithm by \citet[][]{Eisenstein07} with extensive analysis with two-point statistics and BAO fitting. Our findings are
\begin{itemize}
    \item The two algorithms are largely comparable in two-point statistics, which include propagator and cross-correlation coefficient.
    \item HE18 is significantly better at removing RSD than ES3, as shown in quadrupole power spectrum.
    \item Fine tuning smoothing along line-of-sight can lead to reconstruction results slightly closer to the initial condition. However, the contribution of this to BAO fits is negligible.
    \item The two algorithms also produce consistent BAO distance fits.
    \item BAO fitting is robust against the difference in the reconstructed power spectrum produced by different algorithms and reconstruction convention (isotropic vs anisotropic reconstruction).
    \item The most important input to BAO fitting is the smoothing scale.
    \item Adding the second order perturbation theory has a negligible effect to reconstruction. 

\end{itemize}

Although the two methods are largely consistent in various aspects, the HE18 algorithm remains an excellent reconstruction method to be employed in the ongoing and future large-scale structure surveys. Even though the difference is minimal in BAO fitting, HE18 algorithm may potentially have more significant benefits in other areas of analysis, for example, the full-shape analysis to constrain $f\sigma_{8}$, for its much better ability in removing RSDs.

While the scope of this work did not include extending the HE18 method to 
be applied on data with incomplete sky coverage and varying selection functions, 
we observe that these extensions should be relatively straightforward. In particular, 
solving for the displacement field is identical to the standard reconstruction method 
when the second order solution is not included. Therefore, much of the same methodology
\citep[e.g.][]{Padmanabhan12,Burden2015} developed for the ES3 method can be directly applied here. 
Furthermore, since the HE18 method keeps track of the overall density field, it would be 
possible to track where the boundary pixels are advected to, and iteratively refine 
the exact volume of the survey. We defer a detailed implementation of these ideas to 
future work.

\section*{Acknowledgements}

We thank Daniel Eisenstein, Fangzhou Zhu, Naim Karacayli, Uddipan Banik, Martin White, David Alonso, and Ryuichiro Hada for helpful discussions. XC is supported by Future Investigators in
NASA Earth and Space Science and Technology (FINESST) grant
(award \#80NSSC21K2041). NP is supported in part by DOE DE-SC0017660.

\section*{Data Availability}


The Python implementations of both the HE18 and ES3 algorithms, two-point statistics calculations, and BAO fitting codes developed in this work are publicly available at \hyperlink{https://github.com/xinyidotchen/reconstruction}{https://github.com/xinyidotchen/reconstruction} and \hyperlink{https://github.com/xinyidotchen/BAOfitting}{https://github.com/xinyidotchen/BAOfitting}. The statistics produced in the analysis are available upon request. The
\textsc{Quijote} simulations used in this work are publicly available with
details in \citet[][]{Navarro19}.



\bibliographystyle{mnras}
\bibliography{recon_paper} 








\bsp	
\label{lastpage}
\end{document}